\documentclass{jfm}
\usepackage{color}
\usepackage{bbm}
\pdfoutput=1

\begin{document}

\newtheorem{lemma}{Lemma}
\newtheorem{corollary}{Corollary}

\shorttitle{Modelling finger propagation in elasto-rigid channels} 
\shortauthor{J. V. Fontana, A. Juel , N. Bergemann, M. Heil, and A. L. Hazel} 

\title{Modelling finger propagation in elasto-rigid channels}

\author
 {
 Jo\~ ao V. Fontana\aff{1},
  Anne Juel\aff{1},
  Nico Bergemann\aff{2},
  Matthias Heil\aff{2},
  \and
  Andrew L. Hazel\aff{2}\corresp{\email{Andrew.Hazel@manchester.ac.uk}}
  }

\affiliation
{
\aff{1}
Manchester Centre for Nonlinear Dynamics and Department of Physics and Astronomy, University of
Manchester, Oxford Road, Manchester M13 9PL, UK
\aff{2}
Department of Mathematics and Manchester Centre for Nonlinear Dynamics and University of Manchester,
Oxford Road, Manchester M13 9PL, UK
}

\maketitle

\begin{abstract}
 We conduct a theoretical study of a two-phase-fluid-structure interaction problem in which air is driven at constant volume flux into a liquid-filled Hele-Shaw channel whose upper boundary is an elastic sheet. A depth-averaged model in the frame of reference of the advancing air-liquid interface is used to investigate the steady and unsteady interface propagation modes via numerical simulation. In slightly collapsed channels, the steadily-propagating interface adopts a shape that is similar to the classic Saffman--Taylor finger in rigid Hele-Shaw cells. As the level of initial collapse increases the induced gradients in channel depth alter the morphology of the propagating finger and promote a variety of instabilities from tip-splitting to small-scale fingering on the curved interface, in qualitative agreement with experiments. The model has a complex solution structure with a wide range of stable and unstable, steady and time-periodic modes, many of which have similar driving pressures. 
We find good quantitative agreement between our model and the experimental data of Duclou\'{e} \textsl{et al.} (\textsl{J. Fluid Mech. vol. 819, 2017, p 121}) for the finger width, sheet profile and finger pressure, provided that corrections to account for the presence of liquid films on the upper and lower walls of the channel are included in the model.

\end{abstract}



\section{Introduction}
\label{Intro}

The displacement of an interface between two Newtonian fluids driven through a narrow gap bounded by elastic walls is a fundamental two-phase-fluid-structure interaction that occurs in many industrial, geophysical and biological processes \citep{JuelARFM2018}. In the absence of fluid inertia, the behaviour of the interface is determined by the interplay between the interfacial surface tension; the viscosities of the fluids; and the elastic properties of the wall. Although the inertialess equations governing the bulk response of the fluids, the Stokes equations, are linear, nonlinearities arise due to the presence of (i) the interface and (ii) the elastic walls.

Elastic walls are not required to elicit complex behaviour; indeed, the Saffman--Taylor viscous fingering instability in a rigid Hele-Shaw channel, a channel whose width is much greater than its height, \citep{SaffmanTaylor1958} is an exemplar of non-trivial interfacial dynamics.
Precisely because of its fundamental nature and the implications for transport of multi-phase flows and flow in porous media, viscous fingering has been extensively studied, see \cite{Homsy1987}, \cite{Couder2000} and \cite{casademunt2004} for reviews. Moreover, the introduction of non-Newtonian effects \citep{lindner_2002} or fluid inertia \citep{chevalier_2006} does not fundamentally change the fingering and can be accommodated by suitable redefinitions of the control parameter.

More recently, attention has turned to control or suppression of the fingering by varying the flow rate \citep{Li2009,Dias2010,Dias2012}; adjusting the viscosity ratio of the two fluids \citep{Bischofberger2014}; introducing particles \citep{Luo2018}; and  modifying the channel geometry either statically  \citep{Al-Housseiny2012} or dynamically via the introduction of elastic walls \citep{Draga2012PRL}. The majority of these studies have been conducted in radial geometries in which the average interfacial propagation speed decreases with distance from the injection point for a constant injected volume flux. In such geometries, a steadily-propagating state is never possible.

\begin{figure}
\center
\includegraphics[scale=1.3]{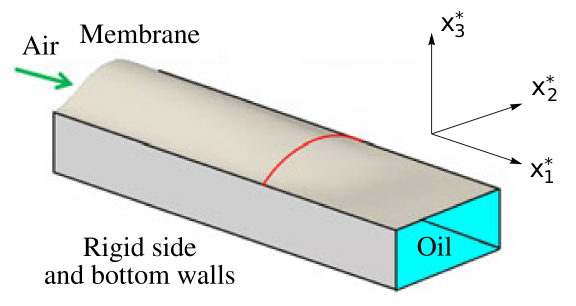}
\includegraphics[scale=1.3]{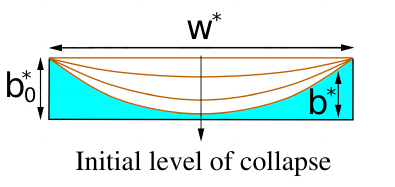}
\caption{The elasto-rigid channel consists of two rigid side walls, a
  rigid lower wall, and a deformable upper wall.
  The cross-section of the channel has width $W^{*}$ and
  undeformed height $b^{*}_{0}$. The
  height of the deformed sheet
  is $b^{*}(x^{*}_{1},x^{*}_{2})$.
  An air finger propagates into the fluid filled channel along the $x^{*}_{1}$ direction.}
\label{Channel_sketch}
\end{figure}

In this paper, we replace the upper wall of an otherwise rigid Hele-Shaw channel by an elastic sheet, see figure \ref{Channel_sketch}, to make an elasto-rigid channel. If fluid is injected from one end of the channel then it is possible for a steadily-propagating state to develop. The propagation of an air finger into a collapsed elasto-rigid channel is a simplified model for pulmonary airway reopening \citep{Gaver1996,HazelHeil2003,Heap2009,Ducloue2017a}, but our focus in this paper is primarily on the nonlinear behaviour of the depth-averaged, elasto-rigid system rather than on any applications to pulmonary mechanics. 

Our study directly complements the experimental investigations of \cite{Ducloue2017a,Ducloue2017b} who observed several different modes of interface propagation in elasto-rigid channels. \cite{Ducloue2017a} found that for constant interfacial propagation speed the complexity of the propagation modes increases with increasing levels of initial collapse of the channel. For modest initial collapse, a single air finger reopens the channel as it propagates steadily and adopts a shape that is reminiscent of a single Saffman--Taylor finger in a rigid channel. At higher levels of collapse the elastic channel reopens over a shorter axial length-scale, for a fixed interfacial propagation speed, and, consequently, the interface propagates into a converging gap.
In this geometric configuration, the interface is unstable  and the instability leads to the formation of small-scale unsteady fingers. \cite{Ducloue2017b} conjectured that the small-scale fingers are analogous to those that develop during peeling of an adhesive strip \citep{Mcewan1966peeling} and those seen in the printer's instability on an interface between two rotating rigid cylinders \citep{Couder2000}. \cite{Callum2020} investigated the behaviour of a strongly collapsed elasto-rigid channel experimentally and found a number of different finger morphologies whose geometric complexity increased with increasing propagation speed from simple  Saffman--Taylor-like fingers to highly disordered interfaces with multiple tips that evolve continuously in time.
\cite{Callum2020} also identified regions of non-trivial transient dynamics that suggested the existence of unstable states mediating the transition between different steadily-propagating states.

 In contrast to the elasto-rigid system, the experimentally observed two-phase flow in an equivalent rigid geometry over the same parameter range is relatively simple. If a more viscous fluid (oil) is displaced by a less viscous one (air) injected from one end of a rigid Hele-Shaw channel, an initially flat interface can exhibit multiple tips transiently, but ultimately a single symmetric finger emerges and propagates at constant speed \citep{SaffmanTaylor1958}. 
In this geometry, the height of the channel's cross-section is much smaller than its width and consequently the flow within the channel can be effectively described using a depth-averaged theory. The simplest two-phase, depth-averaged model does not include surface tension because the perturbation to the interface curvature does not feature at leading order in the expansion in inverse cross-sectional aspect ratio. In the absence of surface tension, however, the model has continuous families of symmetric and asymmetric \citep{SaffmanTaylor1959} solutions at the same flow rate. \cite{McLeanSaffman1981} showed that the \textsl{ad hoc} introduction of surface tension qualitatively reproduces the experimental observations by selecting a single finger from the symmetric solution family at each flow rate. Additional unstable symmetric solutions of the \cite{McLeanSaffman1981} model were later found by \cite{Romero1982} and \cite{VandenBroeck1983} and shown to correspond to symmetric fingers with multiple tips by \cite{mccue2015}.  \cite{ParkHomsy1984} showed that quantitative agreement with experiments requires the inclusion of corrections due to the presence of liquid films that remain on the channel walls after propagation of the air finger. The necessity of including these liquid-film corrections  was later confirmed by the detailed experiments of \cite{Tabelin1986}. 

 Although the depth-averaged model system describing two-phase flow in a rigid Hele-Shaw channel has multiple possible solutions  only the symmetric, single-tipped finger is stable \citep{Tanveer1987,Bensimon1987}. Experimental observations have shown, however, that the finger becomes unstable to tip splitting \citep{Tabelin1986,Tabeling1987} and fluctuations in width \citep{Moore2003} at high driving flow rates in channels of sufficiently large cross-sectional aspect ratio. \cite{Tabeling1987} showed that the flow rate at which instabilities first occur is strongly dependent on channel roughness and
 \cite{Couder2000} suggested that the tip-splitting instability is a noise-induced subcritical transition to nearby alternative states.
 
Rather than relying on uncontrolled perturbations to provoke instability, further studies introduced well-defined perturbations into either the depth-averaged model system or the geometry of the Hele-Shaw channel; see the review by \cite{Couder2000}. These perturbed systems exhibit symmetry breaking and tip splitting, as well as periodic and complex time-dependent behaviour. More recently, \cite{Thompson2014} introduced a prescribed depth perturbation into the model of \cite{McLeanSaffman1981} and showed that this leads to interaction between solutions of the unperturbed system. For example, the symmetric finger exchanges stability with an asymmetric finger at a critical flow rate via a symmetry-breaking bifurcation. The solution structure and sequence of symmetry-breaking and Hopf bifurcations agreed qualitatively with previous experimental observations in channels with cross-sections designed to mimic collapsed elastic tubes \citep{Lozar2009,Pailha2012}. Quantitative agreement between the depth-averaged model and experimental measurements of finger widths for the multiple solutions was subsequently obtained for the the depth-perturbed channels with sufficiently large cross-sectional aspect ratios \citep{FrancoGomez2016}. Thus, in perturbed rigid channels the multiple solutions of the depth-averaged model can be directly related to the complex behaviour observed in experiments. 

Having established that depth-averaged models can be used to describe the observed two-phase flow phenomena in perturbed rigid Hele-Shaw channels, our aims in the present study are twofold: (i) to develop an accurate, depth-averaged model for the elasto-rigid system; and (ii) to use the model to examine the connection between the multiple modes of finger propagation observed by \cite{Ducloue2017a} and the known multiple solutions in depth-averaged models of two-phase flow in perturbed rigid Hele-Shaw channels
\citep{FrancoGomez2016}.

The rest of this paper is divided into three parts. In \S \ref{Model}, we describe the depth-averaged model used to describe the propagating finger and the reopening of the channel as well as its numerical solution using finite element methods. In \S \ref{Results} we demonstrate the good quantitative agreement between the model and the experimental data of \cite{Ducloue2017a} and present illustrative results showing the qualitative behaviour of the model. Finally, in \S \ref{Conclusion} we summarise our findings and describe a dynamic scenario consistent with our results.

\section{Model}
\label{Model}

\begin{figure}
\center
\includegraphics[scale=0.9]{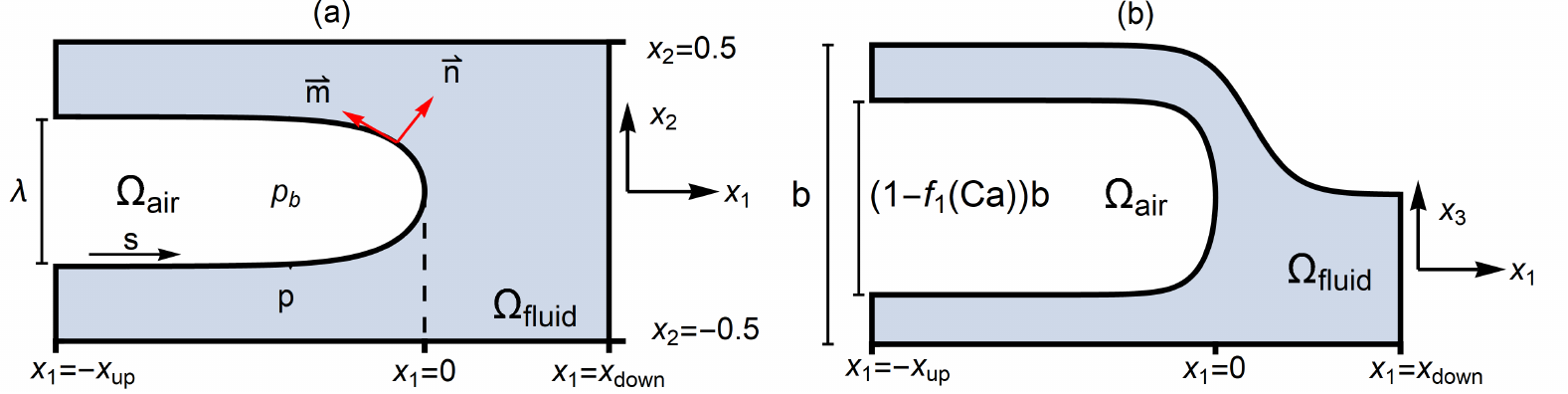}
\caption{(a) Numerical domain in the frame of reference moving with the finger tip: $x_{2}=-0.5$ and $x_{2}=0.5$ are the rigid side walls; $x_{1}=-x_{\mathrm{up}}$ is the upstream end of the domain and $x_{1}=x_{\mathrm{down}}$ is the downstream end of the domain.
(b) Sketch of the thin layers of viscous fluid left behind of the advancing interface, at $x_{2}=0$. The total thickness of the film layers is $f_{1}(Ca)b$. The thickness of the air finger is $(1-f_{1}(Ca))b$.}
\label{Domain}
\end{figure}

We consider the constant-volume-flux 
propagation of an air finger, modelled as an inviscid
fluid at constant pressure, into an elasto-rigid channel containing
an incompressible, Newtonian viscous fluid, see figure
\ref{Domain}. The channel geometry is identical to that used in the
experiments of \cite{Ducloue2017a,Ducloue2017b} and
consists of a rigid base, rigid side walls and a compliant elastic
sheet as the upper boundary. The channel has a width $W^*$ and undeformed
height $b_{0}^*$, with (undeformed) aspect ratio $\alpha \equiv
W^*/b_{0}^* \gg 1$. The elastic sheet has Young's modulus $E^*$,
Poisson's ratio $\nu$ and thickness $h^*$.
The fluid has a dynamic viscosity $\mu^*$ and the constant air-liquid surface
tension is given by $\gamma^*$.
Throughout the paper an asterisk is used to distinguish dimensional quantities
from their non-dimensional equivalents. The initial level of collapse of the
channel, quantified by the channel's cross sectional area $A^{*}_{\infty}$, is set by adjusting the transmural (internal minus external)
pressure, see \S \ref{Fitting}. We choose the external pressure to be our reference pressure and set it to zero.

The modelling framework follows that developed and validated in
studies of radial finger propagation in elastic-walled, Hele-Shaw cells
\citep{Pihler2013, Pihler2014,
  PihlerPeng2015,PengPihler2015,Pihler2018}.
The fluid mechanics is described using depth-averaged,
lubrication equations and the elastic sheet is modelled using F\"{o}ppl--von
K\'{a}rm\'{a}n plate theory, a moderate rotation theory that includes the in-plane stress
contributions to the total force balance. The new features in the
present model, compared to that described by \cite{Pihler2018}, are:
(i) the channel geometry means that the equations are most naturally
formulated in Cartesian, rather than cylindrical polar coordinates;
(ii) the equations are presented in a frame that moves with the tip of
the air finger so that steady
states correspond to steadily-propagating (travelling-wave) solutions; and (iii) the
elastic sheet is horizontally clamped to the side-walls of the channel and is
subject to  an in-plane pre-stress. 

Cartesian coordinates are defined in the frame moving with the instantaneous axial speed of the finger, $u^{*}_{f}(t)$, such that the coordinate $x_{1}^*$ is aligned with the channel axis, $x_{2}^*$ spans the channel width and $x_{3}^*$
is the out-of-plane coordinate (see figure \ref{Domain}). The
notional flow domain is $-\infty<x_{1}^*<\infty$,
$-W^*/2\leq x_{2}^* \leq W^*/2$ and $0 \leq x_{3}^* \leq b^*(x_{1}^*,x_{2}^*)$, where $b^*$ is the distance between the sheet and the bottom wall.

 We non-dimensionalise the in-plane coordinates using the channel width, 
$x^*_{1,2}=W^*x_{1,2}$, and out-of-plane coordinate using the undeformed channel height, $x^*_{3}=b^*_{0}x_{3}$. All three components of the displacement of the elastic sheet are non-dimensionalised using the channel width $W^{*}$,
$(v_{1}^{*}, v_{2}^{*}, w^{*}) = W^{*} (v_{1}, v_{2}, w)$. The flow is driven by the injection of air at a constant flow rate $Q^*$, and we non-dimensionalise the fluid velocity using the  in-plane velocity scale $\mathcal{V}^*=Q^*/(W^*b^*_{0})$. The natural time
scale is thus $\mathcal{T}^*=W^*/\mathcal{V}^*$ and 
the fluid pressure is non-dimensionalised using  $\mathcal{P}^*=12\mu^*\alpha^2/\mathcal{T}^*$. 

After applying the Reynolds lubrication approximation, the governing equation for the fluid pressure, $p$ in the frame moving with instantaneous speed $u_{f}(t) = u_{f}^{*}/\mathcal{V}^{*}$ is
\begin{equation}
\frac{\partial b}{\partial t} - u_{f} \frac{\partial b}{\partial x_{1}}  = b^{3} \frac{\partial^{2} p}{\partial x_\alpha\partial x_{\alpha}},
\label{governing_eq}
\end{equation}
where we use summation convention with Greek indices taking the values $\alpha=1,2$. We determine the unknown speed $u_{f}(t)$ by insisting that the finger tip, defined to be the maximum $x_{1}$ coordinate on the interface, is located at zero, which removes the translational invariance of the system.
The local height of the channel is given by
\begin{equation}
  b(x_{1},x_{2},t) = 1 + \alpha w, \label{eq:channel_h}
\end{equation}
  where $w$ is the dimensionless displacement of the sheet in the $x_{3}$ direction; and the displacement is determined from the F{\"o}ppl--von K\'{a}rm\'{a}n equations \citep{landaulifshitz} in the moving frame
\begin{equation}
\left(\frac{\partial^{2}}{\partial x_{\alpha}\partial
  x_{\alpha}}\right)
\left(\frac{\partial^{2}}{\partial x_{\beta}\partial
  x_{\beta}}\right)w - \eta \frac{\partial}{\partial x_{\beta}} \left( \sigma_{\alpha\beta} \frac{\partial w}{\partial x_{\alpha}} \right) = P, \quad
\frac{\partial \sigma_{\alpha\beta}}{\partial x_{\beta}} = 0,
\label{Fvk}
\end{equation}
where $P$ is the pressure load on the sheet, non-dimensionalised using the bending stiffness
$\frac{E^*}{12(1-\nu^2)} \left( \frac{h^*}{W^*} \right)^3$ 
and the parameter $\eta = 12(1-\nu^2) \left( \frac{W^*}{h^*} \right) ^2$
describes the relative importance of the in-plane and
bending stresses. The components of the in-plane stress tensor, $\sigma_{\alpha\beta}$ are
\begin{equation}
\sigma_{11} = \sigma_{11}^{(0)} + \frac{\left( \epsilon_{11} +\nu \epsilon_{22} \right)}{1-\nu^2},\quad
\sigma_{22} = \sigma_{22}^{(0)} + \frac{\left( \epsilon_{22} +\nu \epsilon_{11} \right)}{1-\nu^2}, \quad
\sigma_{12} = \sigma_{21} = \sigma_{12}^{(0)} + \frac{\epsilon_{12}}{1+\nu},
\label{Stress}
\end{equation}
where $\sigma_{\alpha\beta}^{(0)}$ is the in-plane pre-stress and the in-plane strain is
\begin{equation}
\epsilon_{\alpha\beta} =\frac{1}{2} \left( \frac{\partial v_{\alpha}}{\partial x_{\beta}} + \frac{\partial v_{\beta}}{\partial x_{\alpha}} \right) + \frac{1}{2}\frac{\partial w}{\partial x_{\alpha}}\frac{\partial w}{\partial x_{\beta}}.
\label{strain}
\end{equation}

The equations governing the fluid mechanics (\ref{governing_eq}) and solid mechanics (\ref{Fvk}) are coupled via (i) the displacement of the elastic sheet $w$, which affects the channel height $b$ through equation (\ref{eq:channel_h}); and (ii) the fluid pressure load on the sheet, given by  
\begin{equation}
P = \mathcal{I}p_{b} \ \ \ \ \mbox{in} \ \ \Omega_{air},\quad\quad\quad
P =\mathcal{I}p \ \ \ \ \mbox{in} \ \ \Omega_{fluid}.
\label{p_FSI}
\end{equation}
The fluid-structure interaction parameter
\begin{equation}
\mathcal{I}=\frac{144\mu^* \mathcal{V}^* W^{*2}(1-\nu^2)}{\alpha^2 E^* h^{*3}},
\label{FSI}
\end{equation}
measures the ratio between typical viscous stresses in the fluid and the
stiffness of the elastic sheet. As $\mathcal{I} \to 0$ the sheet becomes rigid and stops interacting with the fluid. 

We impose non-penetration of the fluid on the channel side walls and apply clamped boundary conditions to the elastic sheet:
\begin{equation}
\frac{\partial p}{\partial x_{2}} = 0, \ \ \ \ v_{\alpha}=0, \ \ \ \ \ w=0, \ \ \ \ \frac{\partial w}{\partial x_{2}}=0, \quad\mbox{at}\quad x_{2}= \pm 0.5.
\label{side_wall_bc}
\end{equation}

All disturbances should decay far away from the finger tip, as $x_{1}
\to \pm \infty$. Here we truncate the computational domain at
finite distances behind ($x_{1}=-x_{up}$) and ahead ($x_{1}=x_{down}$) of the
finger, see figure \ref{Domain}. We choose $x_{up}=10$ and
$x_{down}=15$, but we have confirmed that increasing the length of the
domain beyond these values does not alter the results to graphical accuracy.
Following \cite{HazelHeil2003}, we impose
\begin{eqnarray}
v_{\alpha}=0, \ \ \ \frac{\partial w}{\partial x_{1}}=0, \ \ \ \ \frac{\partial p}{\partial x_{1}}=0 \ \ \ \ &\mbox{at}& \ \ x_{1}=-x_{up}, \nonumber \\
v_{\alpha}=0,
\ \ \ \frac{\partial w}{\partial x_{1}}=0, \ \ \ \frac{\partial
  p}{\partial x_{1}}= G   \ \ &\mbox{at}& \ \ x_{1}=x_{down},
\label{truncated_BC}
\end{eqnarray}
and determine the unknown pressure gradient $G$ by imposing the condition that the fluid flux at the truncated downstream boundary is consistent with the level of collapse of the channel far ahead of the finger:
\begin{equation}
\int_{-\frac{1}{2}}^{\frac{1}{2}}  \left( -b^3G -bu_{f} \right)|_{x_{1} = x_{down}}  dx_{2} = -A_{\infty}u_{f}.
\label{Flux_ahead}
\end{equation}
$A_{\infty} = A_{\infty}^{*}/(W^{*} b_{0}^{*})$ specifies the dimensionless initial level of collapse of the channel.

Finally, for the boundary conditions on the air-liquid interface, we use the
same modelling assumptions as \cite{PengPihler2015} and \cite{Pihler2018}, and incorporate the presence of the liquid films into the kinematic and dynamic boundary conditions. The kinematic condition is
\begin{equation}
\left( 1-f_{1}(Ca) \right)\left[\frac{\partial \textbf{R}}{\partial t}
  + u_{f} \mbox{\boldmath$e$}_{1}\right]\cdot \textbf{n} =
-b^{2} \frac{\partial p}{\partial x_{\alpha}} n_{\alpha}\ \ \mbox{on} \ \ \partial \Omega_{air},
\label{K_bc}
\end{equation}
where $\textbf{R}(s,t)$ is the position of the advancing
air-fluid interface in the moving frame, parameterised by the coordinate $s$, and  $\textbf{n}$ is the in-plane outer unit normal vector to the interface, see figure
\ref{Domain}. The dynamic condition is
 \begin{equation}
\Delta p=p|_{\partial \Omega_{air}} - p_{b}=-\frac{u_f}{12\alpha^2 Ca} \left( \kappa + \alpha\frac{2}{b}f_{2}(Ca) \right),
\label{D_bc}
\end{equation}
 where $\kappa$ is the in-plane curvature of the interface and the capillary number $Ca=\mu^* u_{f}^{*}/\gamma^{*}$ is based on the instantaneous velocity of the finger tip $u_{f}^{*}$.
 The functions $f_{1}(Ca)$ and $f_{2}(Ca)$ model the effects of the deposited
liquid films which are directly related to the propagation speed of the
finger, rather than the flow rate. Following \cite{Aussillous_2000}, \cite{PihlerPeng2015} and \cite{PengPihler2015} we take
\begin{eqnarray}
f_{1}(Ca)=\frac{Ca^{2/3}}{0.76+2.16\,Ca^{2/3}}  , \ \ \ f_{2}(Ca)=1+\frac{Ca^{2/3}}{0.26+1.48\,Ca^{2/3}} + 1.59\,Ca.
\label{f1}
\end{eqnarray}
The effects of the liquid films can be neglected by taking $f_{1}(Ca)=0$ and $f_{2}(Ca)=1$.

The governing equations (\ref{governing_eq}) -- (\ref{Fvk}) and boundary conditions (\ref{side_wall_bc}) -- (\ref{D_bc}) were solved using a Galerkin finite element method, implemented in the finite element library, \texttt{oomph-lib} \citep{HeilHazel2006}.
We find steadily-propagating states by setting all time derivatives to zero in the governing equations. Branches of steadily-propagating solutions are found via parameter and arclength continuation. We perform a linear stability analysis of the steadily-propagating solutions at a fixed flowrate $Q$, as opposed to fixed capillary number $Ca$, for consistency with the experiments. In this analysis, we find the eigenvalues, $\lambda$, of the linearised system of equations derived by posing a solution of the form $\mbox{\boldmath$u$} = \mbox{\boldmath$u$}_{ss}(\mbox{\boldmath$x$}) + \epsilon\, \mbox{e}^{\lambda t} \widehat{\mbox{\boldmath$u$}}(\mbox{\boldmath$x$})$ and retaining only terms of $O(\epsilon)$ where $\epsilon \ll 1$. Here, $\mbox{\boldmath$u$}_{ss}$ is the steadily-propagating solution and $\widehat{\mbox{\boldmath$u$}}$ is the associated eigenfunction. Finally, we investigate the nonlinear stability of the steadily-propagating solutions by conducting time simulations of the full system of governing equations.

 The interior of the fluid domain is remeshed at regular intervals in response to a spatial error measure to improve accuracy, and to prevent
excessive mesh distortion. We use a ZZ error
estimator \citep{Zienkiewicz1992} based on the continuity of
$\textbf{U} = -b^2 \nabla p -\textbf{U}_{f}$ between the bulk
elements. In time simulations, the time derivatives were discretised
using a second-order adaptive BDF scheme, where the temporal error was
based on the error estimate for the position of the air-liquid interface.
The resulting set of discrete equations was solved by Newton's method, using the
sparse direct solver SuperLU \citep{Demmei1999} as a linear solver. The number of elements and unknowns varied throughout the simulations, reaching maxima of 15,000 and 200,000, respectively. For linear stability
analysis of the steady states, the solution of the discrete
generalised eigenproblem was obtained via
the Anasazi solver from Trilinos \citep{Heroux2005}.
Further details of the implementation can be found in \cite{Pihler2014,Thompson2014,PihlerPeng2015}.

\section{Results}
\label{Results}

We simulate our system using the same parameters as in the experiments performed by \cite{Ducloue2017a} in which the channel had width $W^*=30$ mm, undeformed height $b_{0}^*=1.05$ mm and length $L^*=60$ cm. The elastic sheet had thickness $h^*=0.34$ mm, Young's modulus $E^*=1.44$~MPa and Poisson ratio $\nu=0.5$. The working fluid was silicone oil with density $\rho^*=973$~kg~m$^{-3}$, dynamic viscosity $\mu^*=0.099$~Pa~s and surface tension $\gamma^*=21$~mN $m^{-1}$. The non-dimensional parameters $\alpha \approx 28.6$, $\eta \approx 70000$ remain fixed, but $Ca$ and $\mathcal{I}$ will vary with the imposed flow rate and $A_{\infty}$ is adjusted to examine the influence of the level of collapse.

\subsection{Initial channel collapse}
\label{Fitting}

\begin{figure}
\center
\includegraphics[scale=0.8]{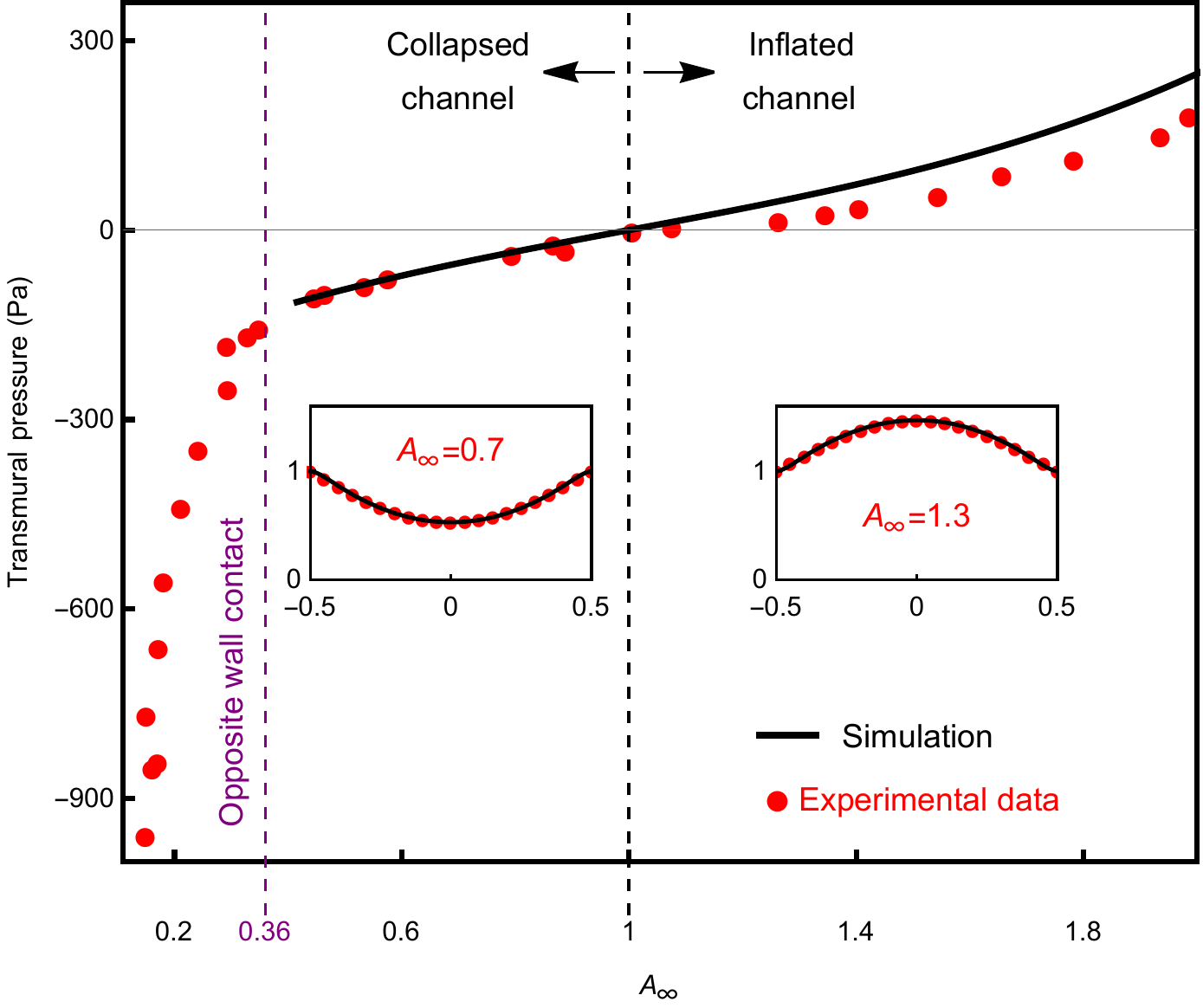}
\caption{Variation of the transmural pressure as a function of the level of collapse, which provides a constitutive relation for the channel (the channel law). The red circles indicate the experiments of \cite{Ducloue2017a}, while the black line is the numerical solution of the F\"oppl--von K\'{a}rm\'{a}n equation (\ref{Fvk}) for experimental parameters and a pre-stress $\boldsymbol{\sigma}_{22}^{(0)*}=30$~kPa and $A_\infty  > 0.36$, the point of near-opposite wall contact.} 
\label{Channel-law}
\end{figure}

We first assess how accurately the F\"oppl--von K\'{a}rm\'{a}n equations capture the deformation of the elastic
sheet in the experimental channel studied by
\cite{Ducloue2017a}. Figure \ref{Channel-law} shows the variation of the transmural pressure (difference between the pressure inside the channel and the
 atmospheric pressure) as a function of $A_\infty$, which represents
 a constitutive relation similar to the so-called tube laws used to model flows in collapsible tubes \citep{Shapiro1977}. We shall refer to the relationship for our system as the channel law. The symbols correspond to the measurements of \cite{Ducloue2017a}. When the transmural pressure is zero, $A_\infty=1$ and the elastic sheet is undeformed. Inset sheet profiles for $A_\infty<1$ and for
 $A_\infty >1$ provide examples of collapsed and inflated channel
 cross-sections, respectively. For $A_\infty \le 0.36$, the
 deformation of the elastic sheet is sufficient for the sheet to come
 into contact with the bottom boundary of the empty channel. 
In this paper, we shall not address the contact problem and instead focus on moderately collapsed/inflated
 channel cross-sections in the range $0.4 \le A_\infty \le 1.2$.

In the experiment, a non-zero pre-stress,
$\sigma_{22}^{(0)*}$, was imposed by hanging evenly distributed weights from
one long edge of the elastic sheet prior to clamping it to the channel
wall. The exact pre-stress imposed was influenced by details of the clamping procedure and was difficult to determine accurately. Hence, in the model we treat the pre-stress as a fitting parameter chosen to achieve the best quantitative match to the experimental results for $0.4
 \le A_\infty \le 1$. The solid line in figure \ref{Channel-law}
 corresponds to the numerical solution of the F{\"o}ppl-von K\'{a}rm\'{a}n
 equations at best fit --- a pre-stress of $\sigma_{22}^{(0)*}=30$~kPa,
 $\sigma_{11}^{(0)*}=0$ and
 $\sigma_{12}^{(0)*}=0$.
 The quantitative
agreement between
 model and experiment over this parameter range extends to the sheet
 profiles shown as insets in figure \ref{Channel-law} for
 $A_\infty=0.7$ and $1.3$, respectively. The sensitivity
 of the channel law to variations in pre-stress was assessed by varying
 $\sigma_{22}^{(0)*}$ by $\pm 2$~kPa (6.6\%), which
 resulted in a variation in the transmural pressure of $\pm 5.7\%$ at
 $A_\infty=0.6$.

 Under perfect clamping
 conditions the system will be up-down symmetric in the sense that the
 same deflection would result from the same transmural pressure
 magnitude irrespective of direction. Imperfections in the
 experimental clamping procedure break the up-down symmetry, but are
 not included in the theoretical model. We
 choose to match the experimental and numerical channel laws for
 collapsed channels ($0.4 < A_\infty \le 1$) rather than for inflated
 channels ($A_\infty > 1$) to ensure that the transmural pressures
 required to set a given level of initial collapse are the same in the
 experiments and the model. Moreover,  the reopening dynamics in the fully
 coupled system occur near the tip of the propagating finger, where
 the channel is typically collapsed. We will
 show in \S \ref{exp_comp} that the remaining discrepancy between experimental
 and numerical channel laws leads to a modest underestimation of the
 inflation far behind the finger tip 
 (see figure \ref{Profile}), but does not appear to
 affect any of the other dynamics.

\subsection{Steady finger propagation}
\label{Steady}

\subsubsection{Comparison with the experiments of \cite{Ducloue2017a}}
\label{exp_comp}

\begin{figure}
	\center \includegraphics[scale=0.35]{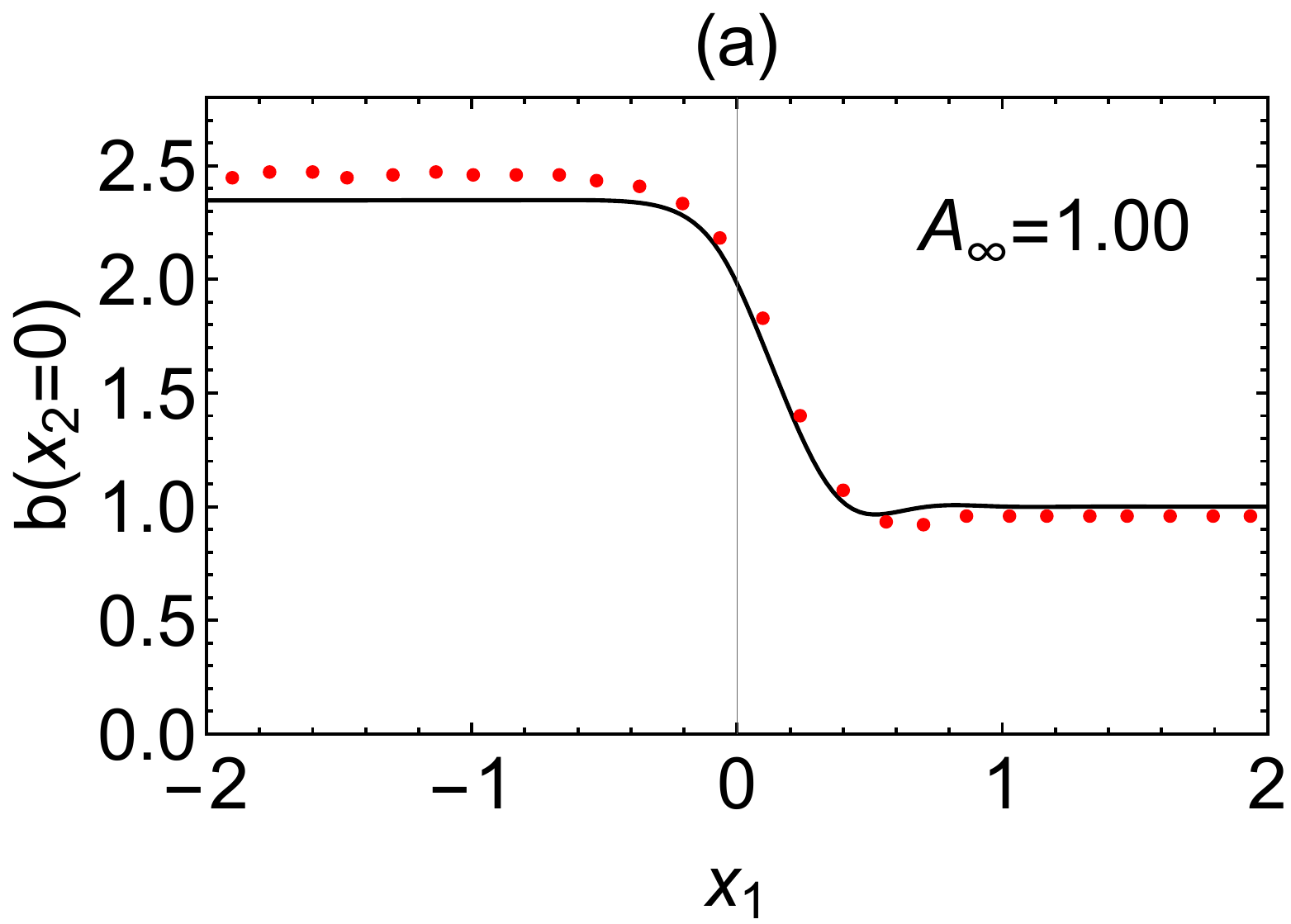}
        \includegraphics[scale=0.35]{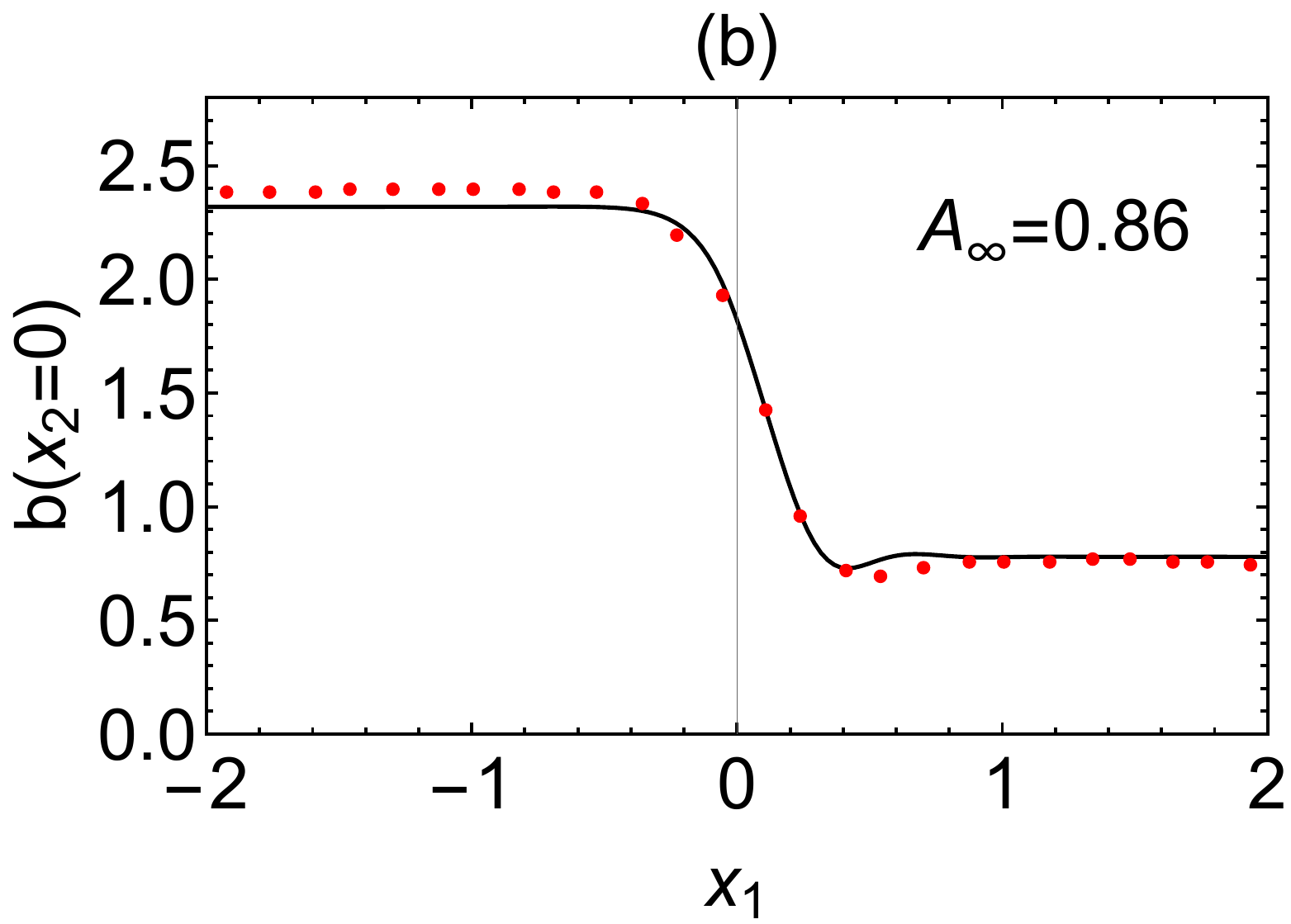}
        \includegraphics[scale=0.35]{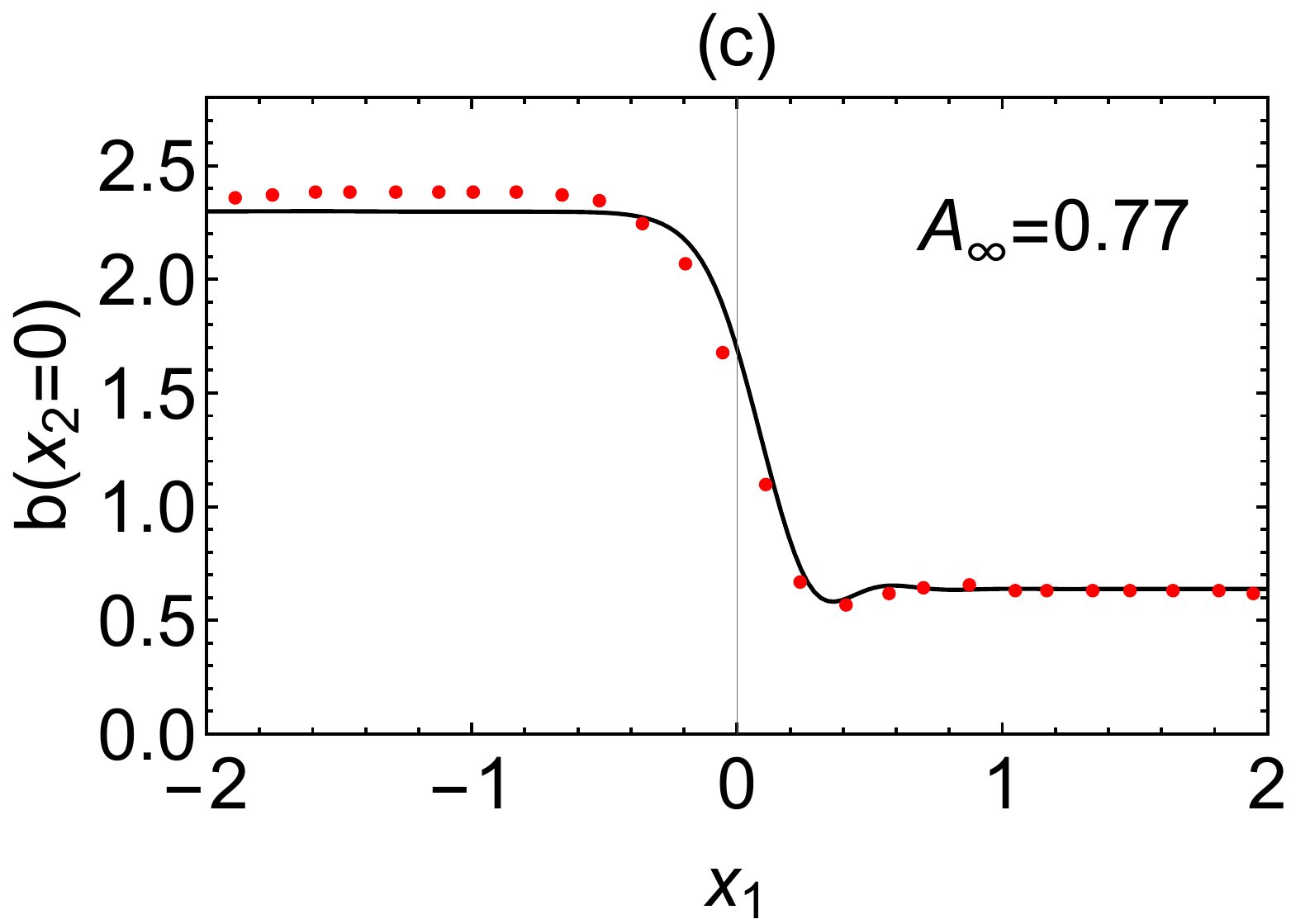}
        \includegraphics[scale=0.35]{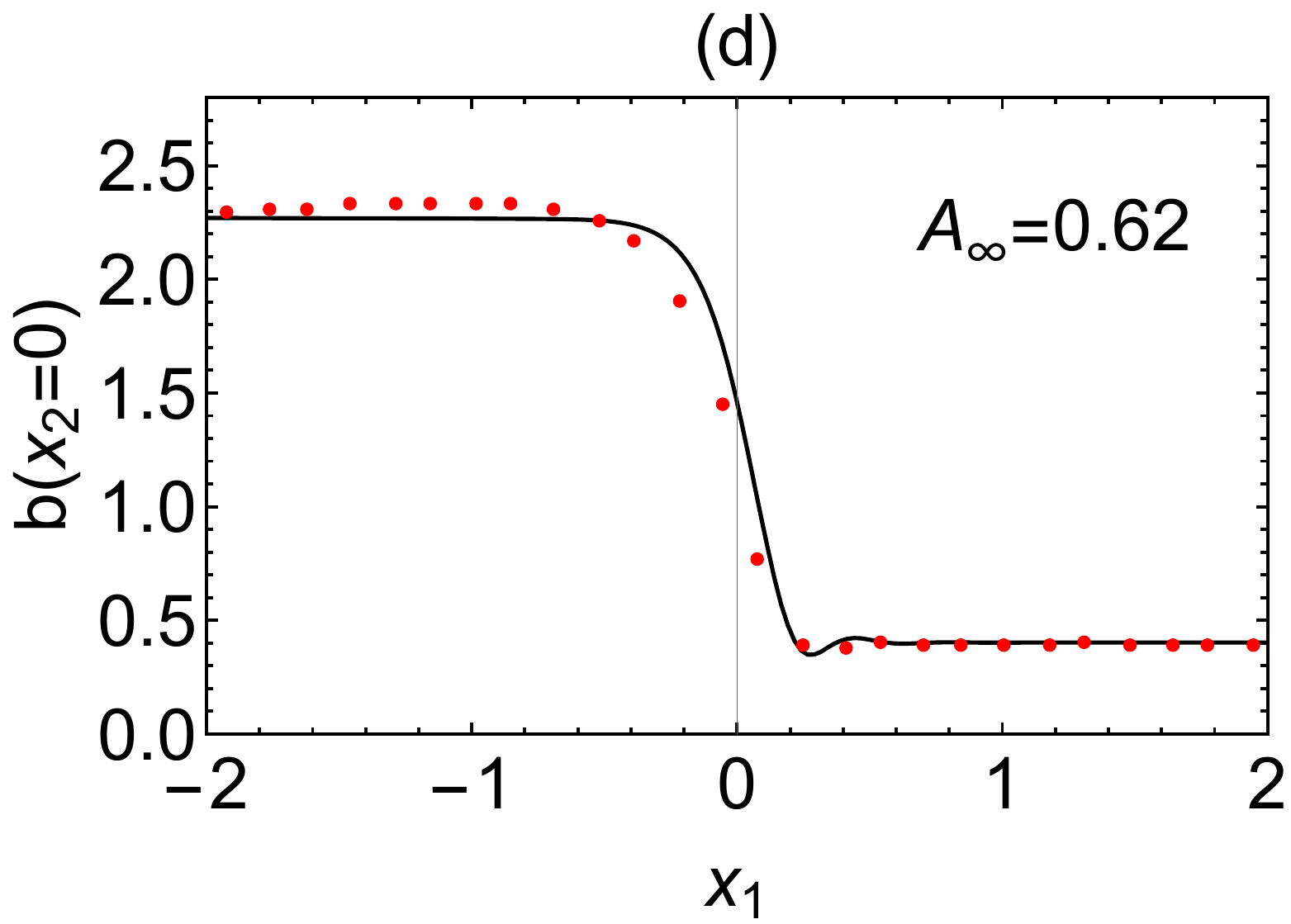}
        \includegraphics[scale=0.35]{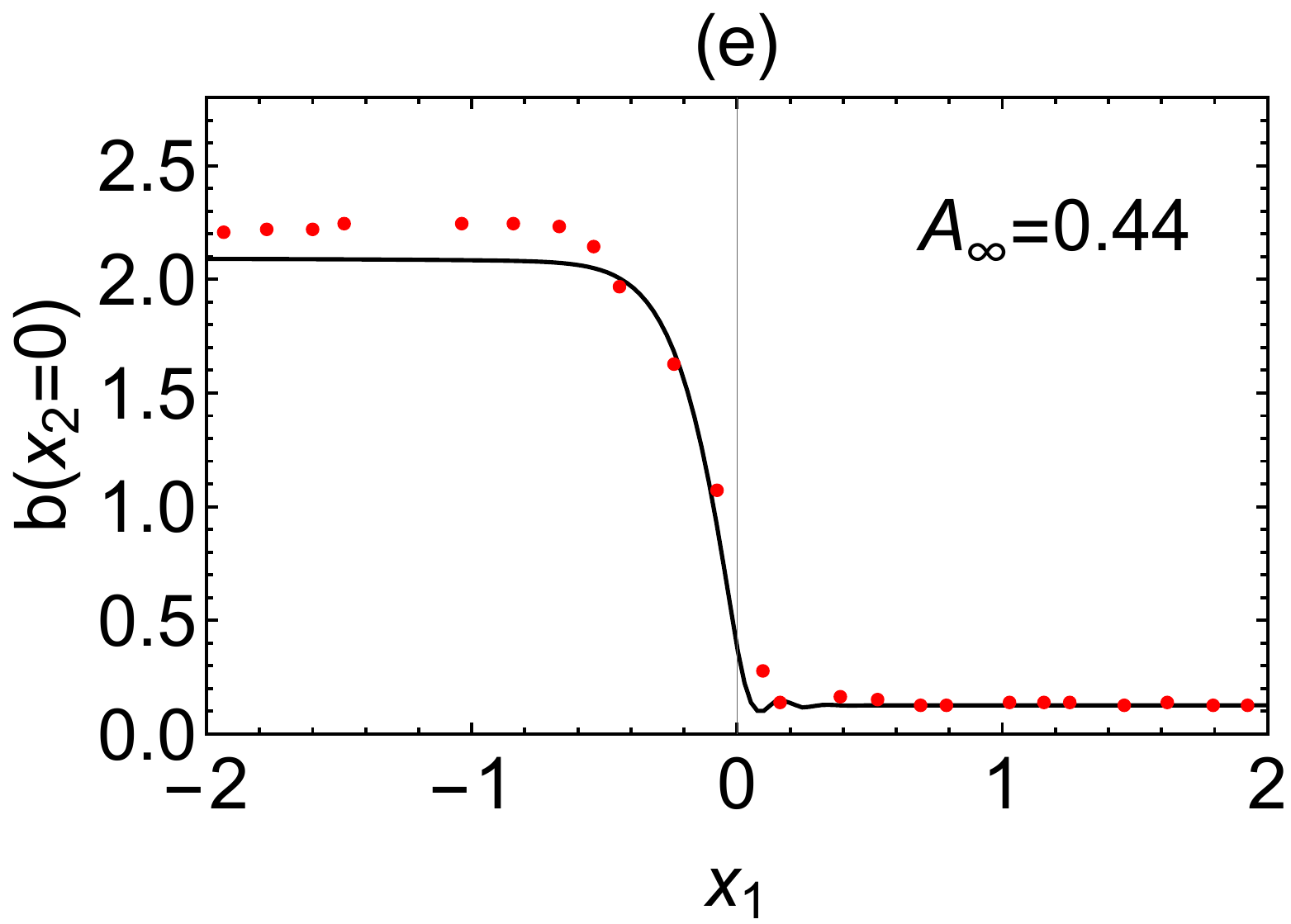}
	\caption{Membrane height along the centreline of the channel
          ($x_{2}=0$) for decreasing values of $A_\infty$. The red
          circles indicate the experiments of \cite{Ducloue2017a},
          while black lines indicate steady numerical
          solutions of the fully coupled fluid-structure interaction
          model. The tip of the air finger is located at $x_{1}=0$.}
	\label{Profile}
\end{figure}

\begin{figure}
\center \includegraphics[scale=0.185]{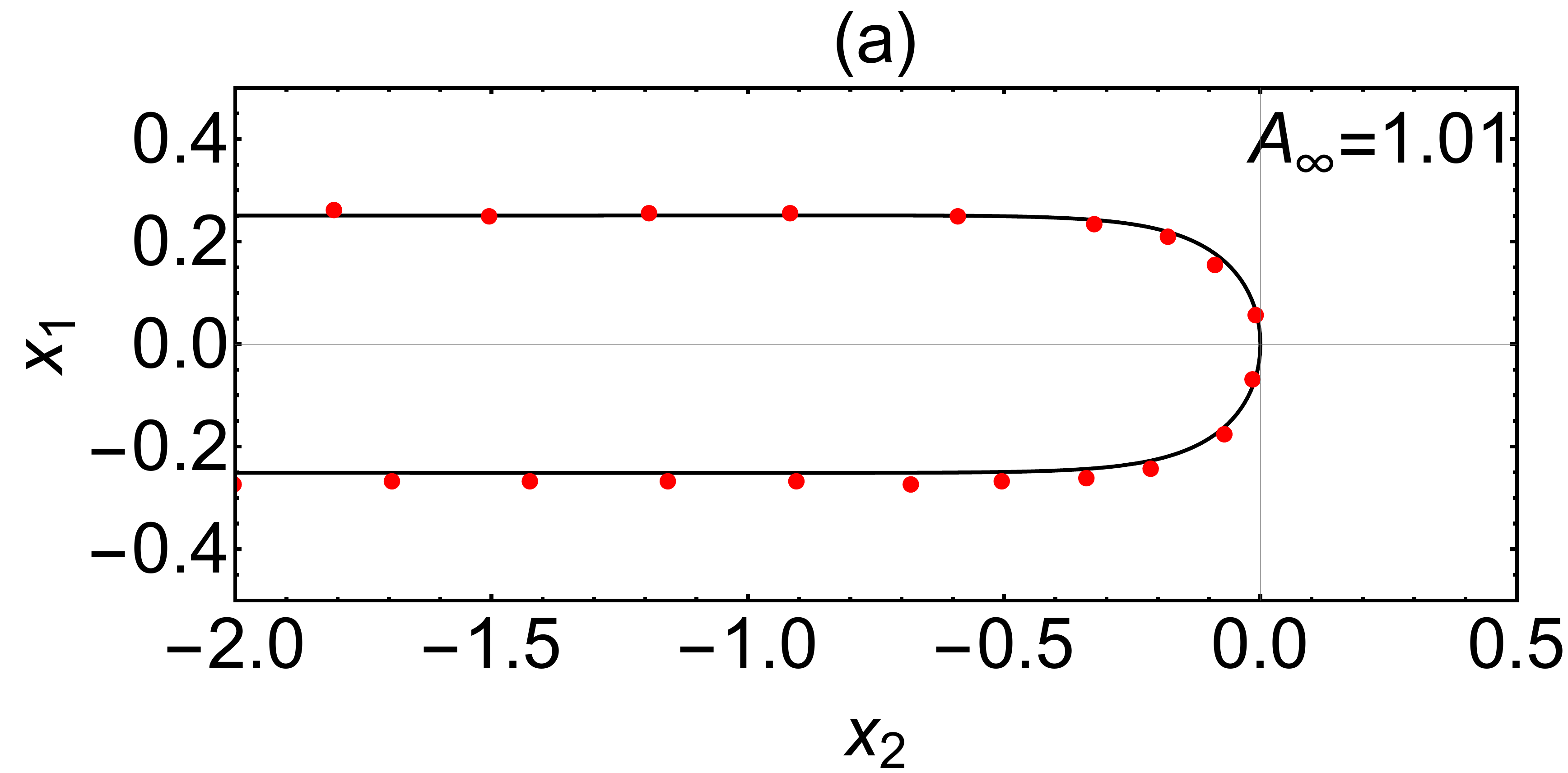}
\includegraphics[scale=0.185]{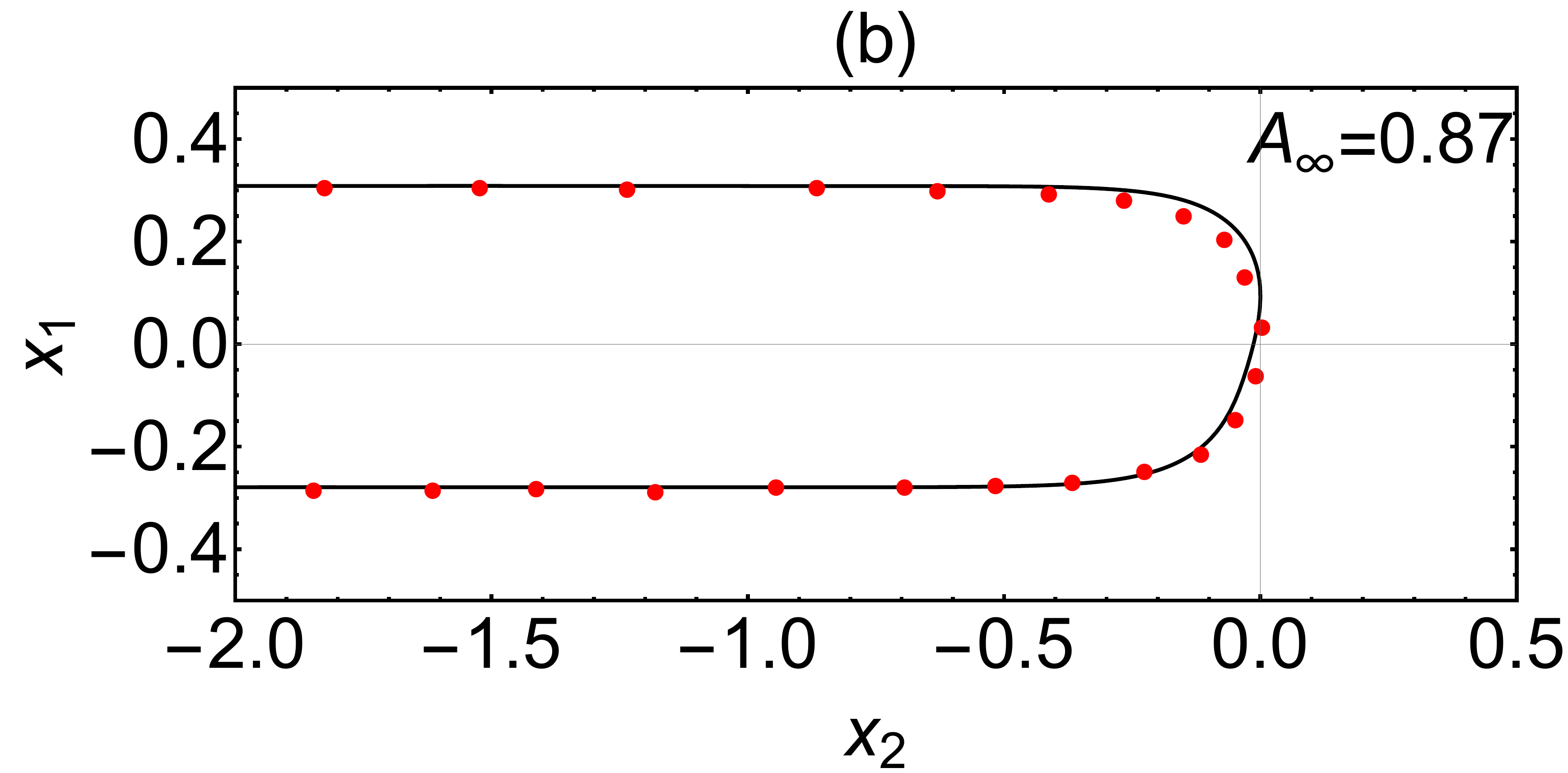}
\includegraphics[scale=0.185]{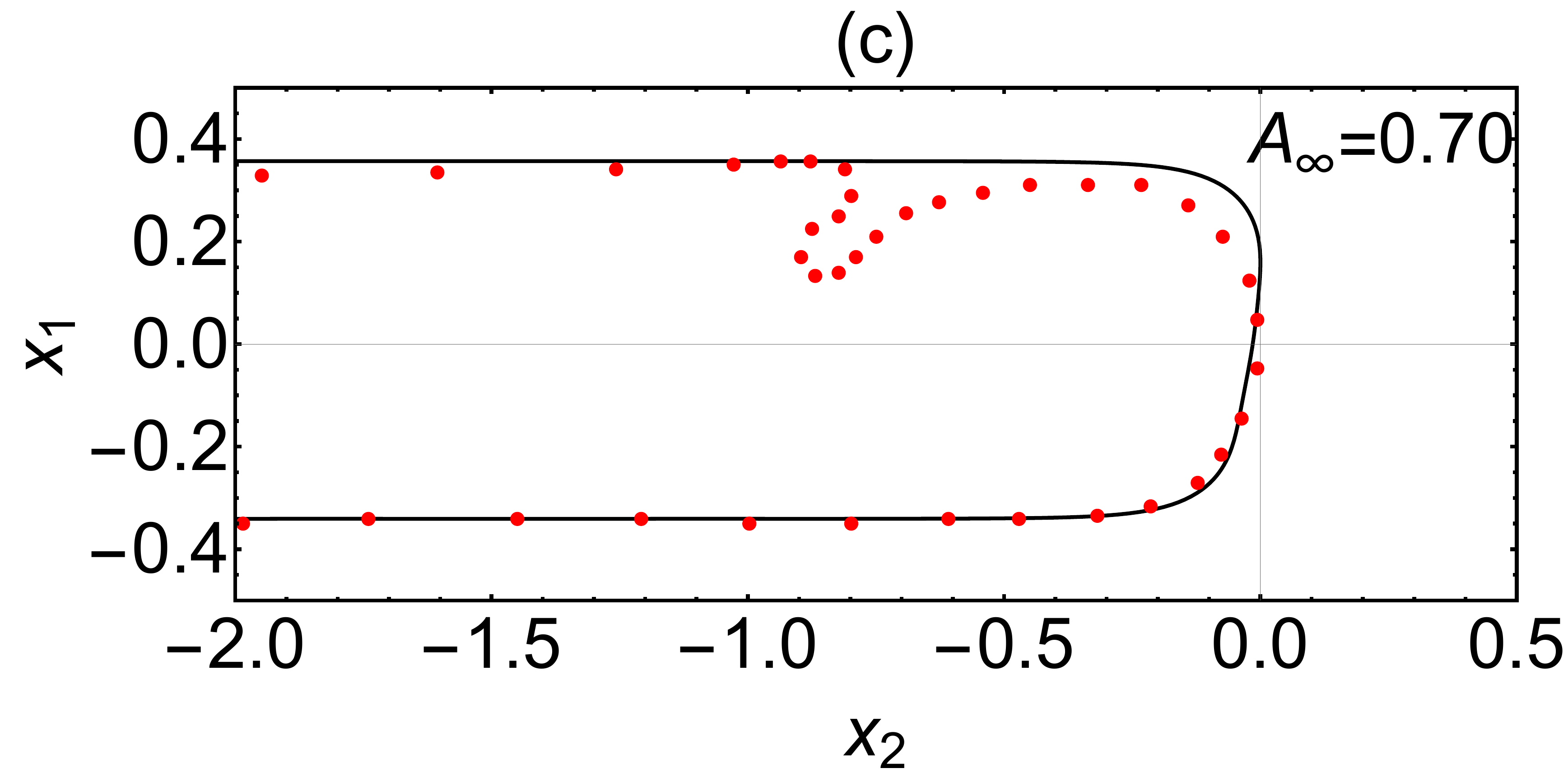}
\includegraphics[scale=0.185]{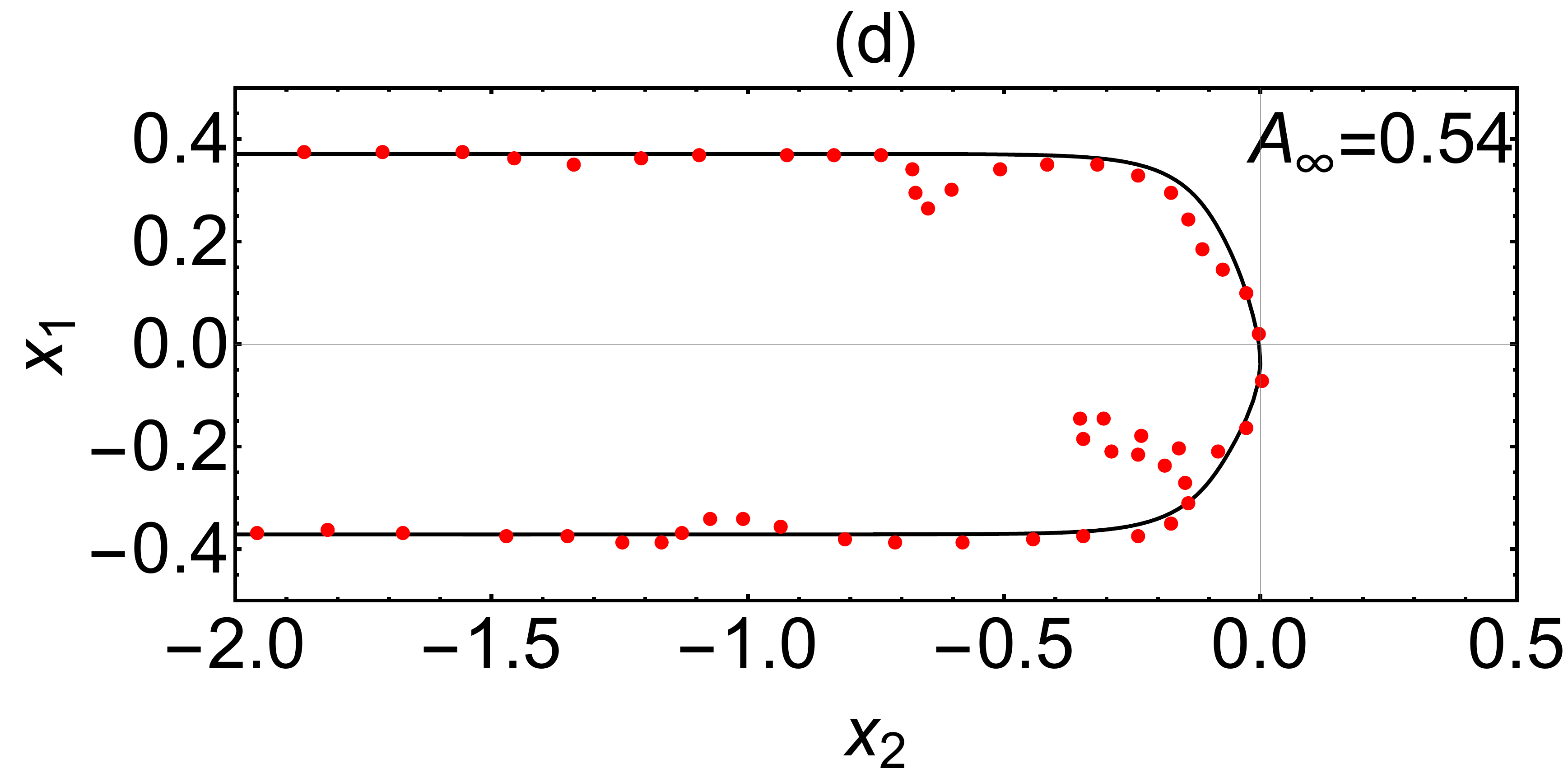}
\includegraphics[scale=0.185]{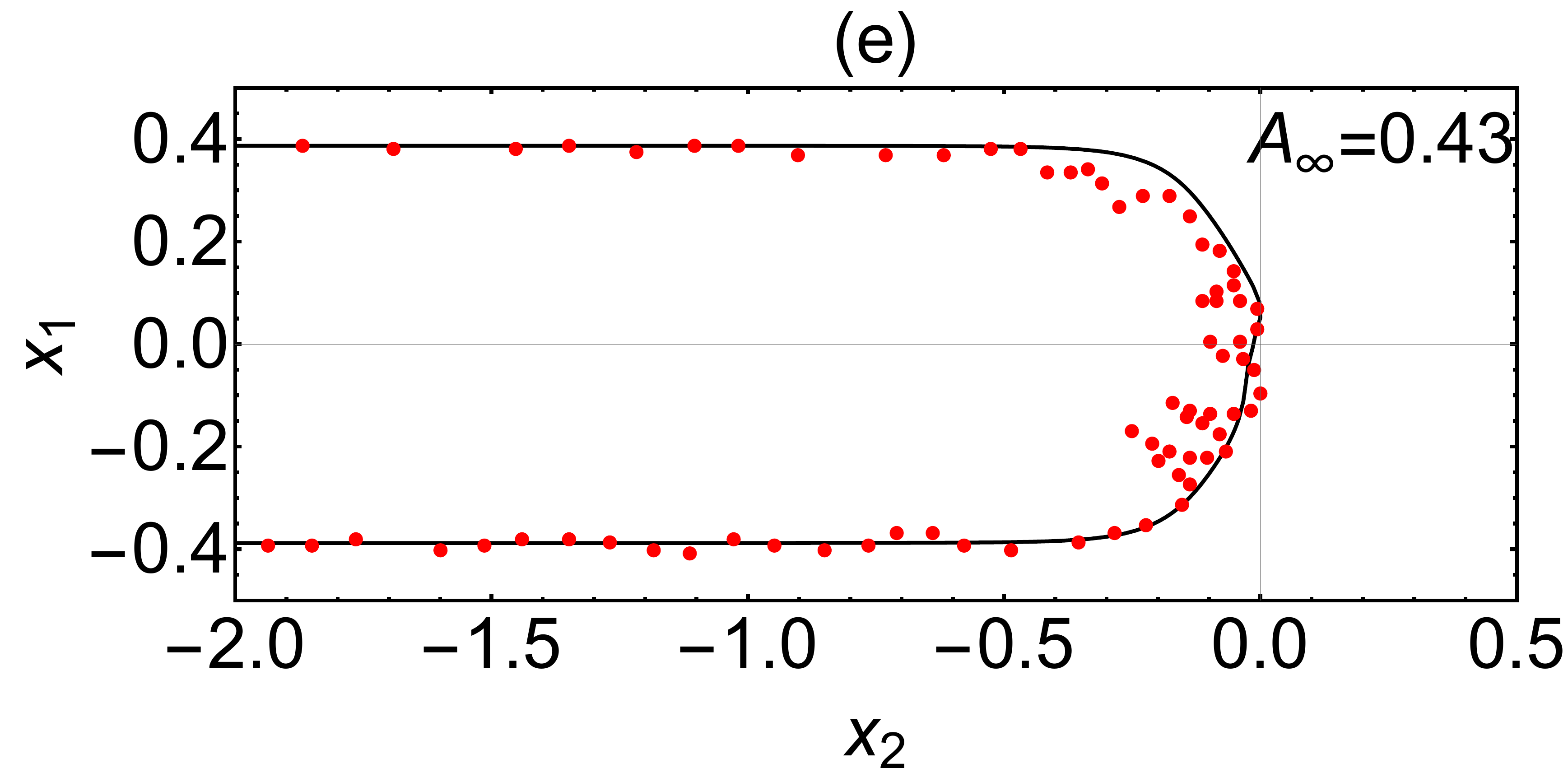}
\caption{Finger shapes delimited by the air-liquid interface for
  decreasing values for $A_\infty$. The red circles indicate the
  experiments of \cite{Ducloue2017a}, while black lines
  show steady numerical solutions of the fully coupled
  fluid-structure interaction model. The experimental fingers shown in (c,d,e)
  are snapshots of unsteady modes of propagation where small scale
  fingers are continually formed near the tip and advected around
  the curved front.}
\label{Interface}
\end{figure}

We examine steady finger propagation
for different levels of initial collapse and a fixed propagation speed
corresponding to $Ca=0.47$.
We present direct comparisons between our numerical calculations 
and the experimental results presented in \cite{Ducloue2017a}. 
Profiles of the elastic sheet measured
along the centreline of the channel at $x_2=0$ are shown in figure
\ref{Profile} and finger shapes viewed from above are shown in figure
\ref{Interface}. Experimental measurements are plotted with red
symbols, while black lines denote the numerical results. The finger tip is located
at $x_{1}=0$ in all the plots shown and was used as the reference point
to align the experimental and numerical results. 

Figure \ref{Profile} shows that as
$A_\infty$ decreases (\textsl{i.e.} the initial level of collapse increases),
the profile steepens in the reopening region and the finger tip ($x_{1} = 0$) is
displaced towards the most collapsed region so that the volume of
fluid ahead of the interface is reduced to a small wedge. These
changes in the channel geometry near the finger tip are associated
with a gradual reduction of the importance of viscous stresses
relative to elastic stresses resulting in the development of 
an elastic peeling mode 
\citep{GaverEtAl1990,Gaver1996} as $A_\infty$ decreases, see also \cite{PengPihler2015}, \cite{PengLister2019} and \cite{Callum2020}.

Figure \ref{Interface} shows that when  $A_\infty=1.01$, corresponding to a slightly inflated sheet, 
the finger propagates steadily and is symmetric about the
centreline $x_2=0$, resembling the classical Saffman-Taylor finger in
a rigid Hele-Shaw channel.

As the initial level of collapse of the channel is increased, the
finger widens and the in-plane curvature of the finger tip
decreases. \cite{Ducloue2017a} found that the finger width increases
linearly with decreasing $A_\infty$, which they explained 
using a simple mass conservation argument. In all cases, the width of
the finger behind the tip predicted by the model 
agrees with the experimental results and, therefore,
obeys the same linear scaling with $A_\infty$.

At $A_\infty =0.87$, the computed finger shape 
has lost its symmetry about the centreline $x_2=0$ to a
finger with a slightly asymmetric tip. This asymmetry is enhanced for
$A_\infty =0.7$, where it can also be seen in the
experimental finger shape and remains at $A_{\infty} =
0.54$ and  $A_{\infty} = 0.43$. The relatively modest
changes of the tip shape for these asymmetric fingers means that their
existence could not be convincingly established from the experimental
data alone and hence they were not identified by \cite{Ducloue2017a,Ducloue2017b}.

The most striking feature of the comparison shown in 
figure \ref{Interface} is that steadily-propagating fingers are not observed in the experiments for
$A_{\infty} \le 0.7$. Snapshots of the observed
small-scale fingering instabilities are presented in these cases.  
The overall shape of the finger is captured accurately by our
steadily-propagating numerical simulations, however, which suggests that, as conjectured by \cite{Ducloue2017a}, fingering instabilities develop on unstable steadily-propagating base states for $A_\infty \le 0.7$. We shall discuss these instabilities further when we present unsteady numerical solutions in \S \ref{unsteady}.
 
\begin{figure}
\center \includegraphics[scale=1.1]{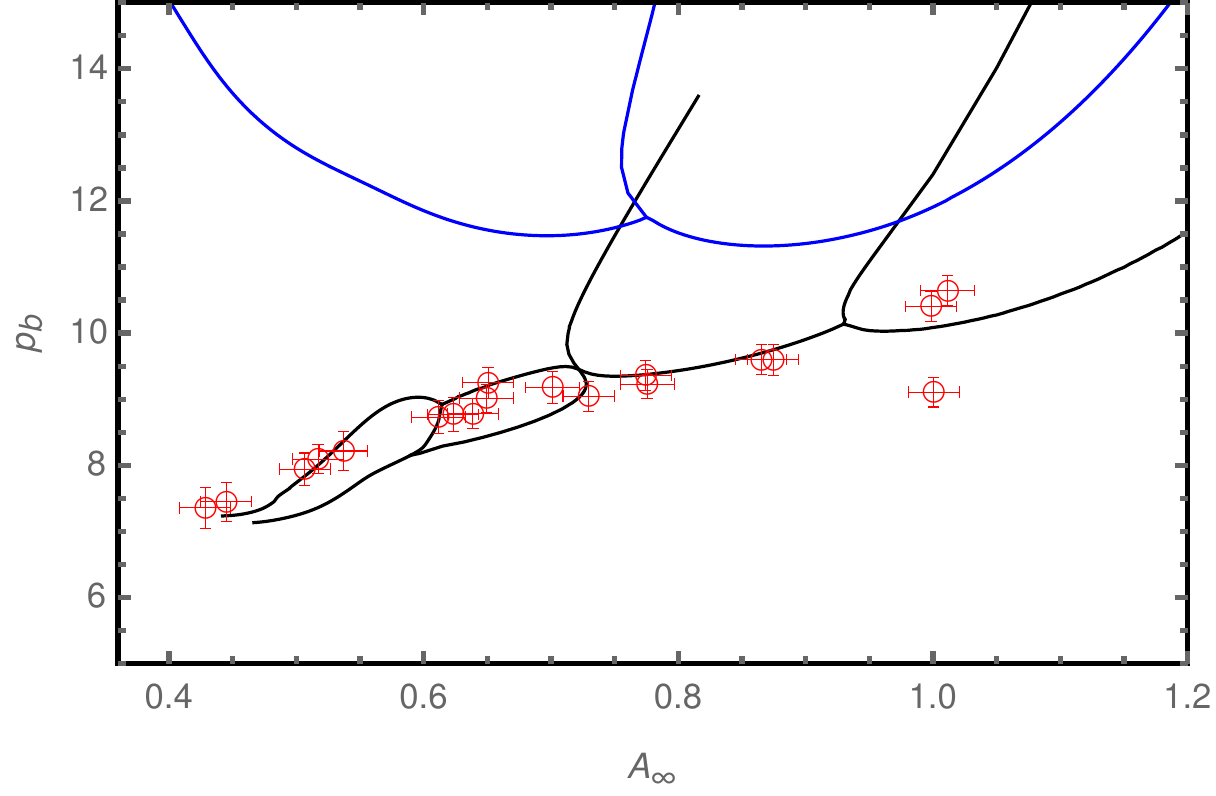}
\caption{Finger pressure $\tilde{p}_{b} = p_{b}^{*}/(\gamma^{*}/b^{*}_{0})$ on the capillary scale as a function of the initial level
  of collapse $A_{\infty}$ for a fixed capillary number of
  $Ca=0.47$. The experimental data from \cite{Ducloue2017a} are shown
  with red circles and the error bars correspond to the standard
  deviations of three experiments. The black lines show
  steady solutions of the model, and
  blue lines are the corresponding solutions without any liquid-film corrections.}
\label{Exp_thin}
\end{figure}

 In figure \ref{Exp_thin}, we present a global measure of the system
 behaviour by plotting the finger pressure as a function of $A_\infty$ at   $Ca=0.47$. The experimental data of \cite{Ducloue2017a}
 are shown with red symbols and the error bars denote the
 standard deviations of three experiments conducted for the same level of  initial collapse and the same flow rate. The black lines in figure \ref{Exp_thin} are steadily-propagating solutions of the theoretical model.
 As in rigid channels, we find that the model has a complex solution structure with multiple steady solution branches connected via bifurcations, which we describe in \S \ref{Multiple_modes}. The experimental measurements are all close, usually to within experimental error, to branches of steadily-propagating numerical solutions. Thus, the model provides a reasonable prediction of the finger pressure observed in the experiment for $0.4 < A_\infty < 1$.

 In addition, the blue lines in figure \ref{Exp_thin} show the results of the model without the liquid-film corrections.  The necessity for the liquid-film corrections to achieve quantitative agreement with experiments in rigid Hele-Shaw cells \citep{Tabeling1987} and elastic cells \citep{PihlerPeng2015,PengPihler2015} has been previously established for individual solutions. We are unaware of studies that have investigated the influence of the liquid films in situations where there are multiple solutions. 
Over the range of $A_{\infty}$ shown in Figure \ref{Exp_thin}, we find that the inclusion of liquid-film corrections has a dramatic effect on that solution structure: the number of solutions at a given value of $A_{\infty}$ differs between the two models for most of the range shown and although both models have a single solution when $0.78 < A_{\infty} < 0.93$ the solutions have different symmetries: the solution in the absence of liquid-film corrections is symmetric about the channel's centreline, whereas it is asymmetric when liquid-films are included, see figure \ref{Pb_Ainf}.  We conclude that the liquid-film corrections are required for both quantitative and qualitative agreement with experimental data.

\subsubsection{Stability analysis and steadily-propagating solution structure}
\label{Multiple_modes}

\begin{figure}
  \center \includegraphics[scale=0.3]{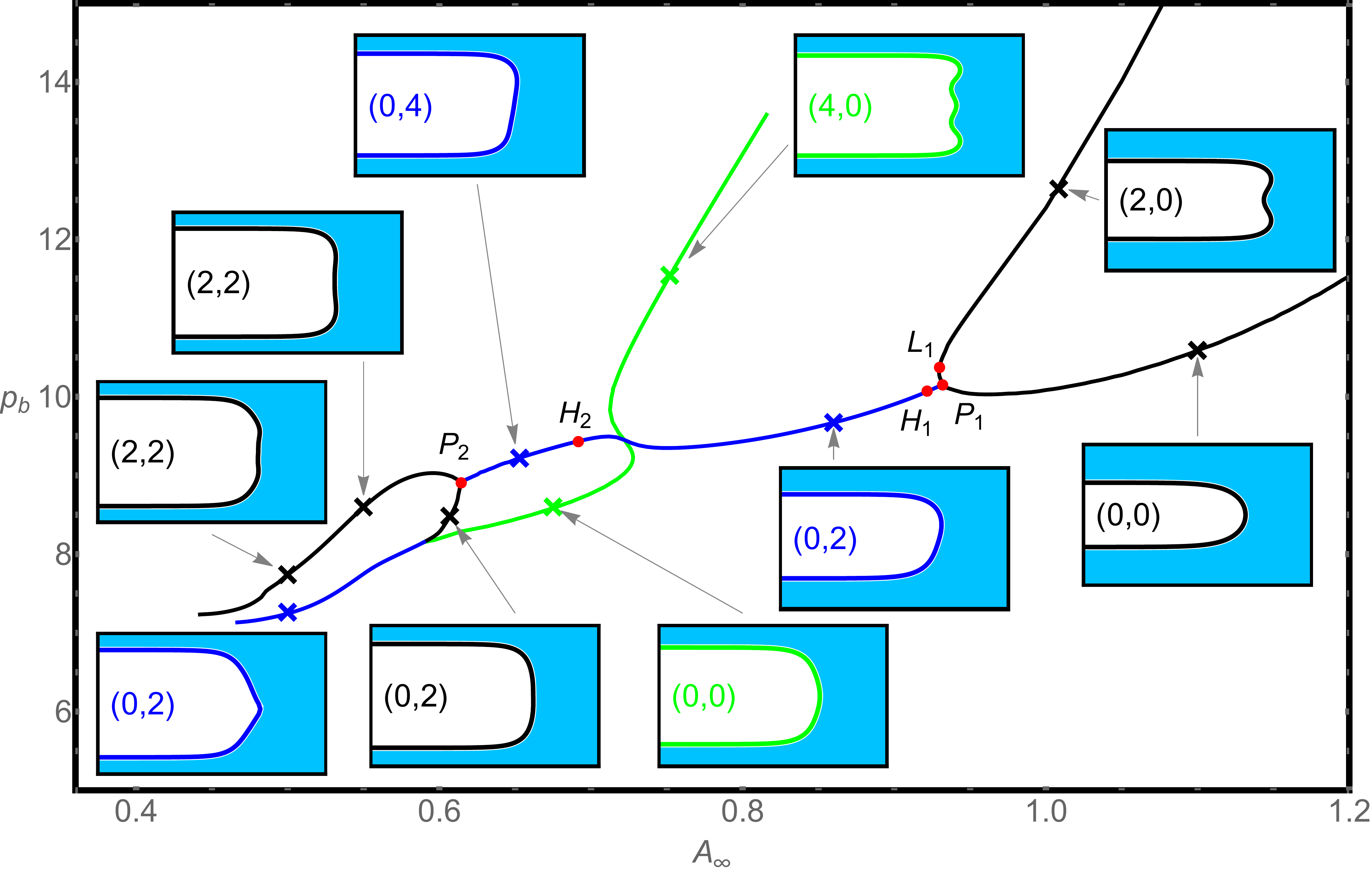}
\caption{Steady numerical solutions with liquid-film corrections shown
  in terms of the variation of the finger pressure $p_{b}$ as a
  function of the initial level of collapse $A_{\infty}$, for a fixed
  capillary number of $Ca=0.47$. The results are similar to those
  shown in figure \ref{Exp_thin}, but here, each solution branch is
  shown with a different colour and the finger morphologies for
  different parameter values are illustrated with inset images.
  Solutions that are symmetric about the channel's centreline are represented as black or green lines; asymmetric solutions are shown as blue lines. $P_1$
  and $P_2$ denote pitchfork bifurcations, $H_{1}$ and $H_{2}$ the
  locus of Hopf bifurcations, and $L_{1}$ is a limit point. The number
  of positive eigenvalues are indicated by the pair of number in the
  inset images, where the first number counts the real positive
  eigenvalues and the second one the number of complex eigenvalues with    positive real part; these occur in complex conjugate pairs}
\label{Pb_Ainf}
\end{figure} 

In figure \ref{Pb_Ainf}, we replot the data for the steady numerical
solutions previously shown in figure \ref{Exp_thin}, but add information about linear stability of the solutions and the location of bifurcations. We 
use a different colour for each solution branch: solutions that are symmetric about the channel's centreline are shown in black or green and asymmetric solutions are shown in blue.
The points $P_1$ and $P_2$ indicate pitchfork bifurcations at which the symmetric solution exchanges stability with a pair of asymmetric solutions.
The points $H_{1}$ and $H_{2}$ are Hopf
bifurcations at which the steadily-propagating solutions become unstable to oscillatory solutions, in the moving frame, and $L_{1}$ is a limit point. Finger morphologies corresponding to each branch are illustrated with inset images. 
The structure becomes increasingly intricate for decreasing values
of $A_\infty$, corresponding to increasing levels of collapse.

The steady solutions in figures \ref{Exp_thin} and \ref{Pb_Ainf} are shown at fixed $Ca$ for comparison with the experimental data. In any given experiment, however, the flowrate, $Q$, is fixed and the finger speed, represented by $Ca$, is free to vary. Hence, we fix the flowrate in our stability analysis, as described in \S \ref{Model}. The linear stability of the solution branches shown in figure \ref{Pb_Ainf} is indicated by the pair 
$(i,j)$, where $i$ and $j$ denote the number of positive
(unstable) real eigenvalues and complex eigenvalues with a positive real part,
respectively. 

For $A_{\infty}>1$, there is a single stable solution (black branch),
that is steady and similar to a Saffman--Taylor finger in a rigid
channel, as previously discussed in \S \ref{exp_comp}. This symmetric
finger persists as $A_\infty$ decreases until it exchanges stability with a
stable asymmetric finger (blue branch) at a supercritical pitchfork
bifurcation $P_1$ ($A_{\infty} (P_1)=0.93$). The resulting unstable
symmetric finger has a nearby limit point $L_{1}$ ($A_{\infty}
(L_{1})=0.927$), beyond which the solution becomes doubly unstable. 
The finger then develops
a region of negative curvature at its tip as $A_\infty$ is
increased. The resulting finger morphology is reminiscent of the first family of Romero--Vanden-Broeck
(RVB) solutions in a rigid Hele-Shaw channel
\citep{Romero1982,VandenBroeck1983,mccue2015,green2017effect}. 
The same solution structure has been observed in rigid channels with depth perturbations \citep{FrancoGomez2016}, in which case the first family of RVB solutions was shown to connect to the Saffman--Taylor finger as the height of the perturbation increased.

As $A_\infty$ decreases from $A_{\infty}(P_1) = 0.93$, the stable
steady mode of finger propagation is asymmetric about the centreline
$x_2=0$ (blue branch), consistent with the experiments shown in \S
\ref{exp_comp}. This finger loses stability to a time-periodic
solution at a Hopf bifurcation $H_{1}$ ($A_\infty (H_{1})=
0.922$), which we show to be subcritical in \S \ref{unsteady}.
A second Hopf bifurcation occurs on the asymmetric branch at ($A_{\infty}(H_{2})=0.69$). We shall discuss the oscillatory modes that emerge from these Hopf bifurcations in \S \ref{unsteady}. 

As the level of collapse increases yet further, the steadily-propagating asymmetric finger (blue branch) regains symmetry about the centreline of the channel at a second pitchfork bifurcation $P_2$ ($A_\infty (P_2)= 0.61$). A linear stability analysis of the branches at some distance from $P_{2}$ is consistent with the local structure being identical to that near $P_{1}$. In other words,  we expect there to be a limit point and a Hopf bifurcation in the vicinity of $P_{2}$, but the details of this region are difficult to resolve.

The symmetric green branch is not connected to any other branches in
the parameter regions that we have examined, despite the fact that it
has the same finger pressure as other solutions for particular
values of $A_{\infty}$. The finger morphology on
the green branch is reminiscent of the second family of RVB solutions
(triple-tipped) for values of $A_\infty \simeq 0.8$, which are
disconnected from the Saffman--Taylor solution branch in a rigid Hele-Shaw channel containing a centered obstacle \citep{FrancoGomez2016}.
For high levels of collapse, the distinction
between the different branches is primarily in the shape of the tip;
the finger widths are approximately equal. For solutions with
equal finger widths moving at the same speed, 
conservation of mass demands that cross-sectional
areas of the channel containing the finger must be equal and hence the
finger pressures of the solutions must also be equal. It is,
therefore, very difficult to distinguish between different solutions
in this region and we suspect that there may be other solutions that
we have not identified. For this reason, the details of the region where the blue, black and green branches appear to meet at $A_{\infty} \approx 0.6$ have not been resolved.

\subsection{Unsteady finger propagation}
\label{unsteady}

In this section we replicate individual experiments by performing time-dependent simulations at fixed flow rates. We complement the linear stability analysis presented in \S \ref{Multiple_modes} by assessing the sensitivity of the steadily-propagating solutions to general perturbations for increasing levels of collapse, $A_{\infty} = 1$, $0.926$, $0.8$, $0.66$, $0.5$, and $0.44$. According to bifurcation diagram shown in figure \ref{Pb_Ainf},
the system should exhibit different dynamics at each of the chosen levels of collapse. Note that these levels of collapse do not correspond directly to the experimental levels of collapse chosen by \cite{Ducloue2017a}.

\begin{figure}
\center
\includegraphics[scale=0.5]{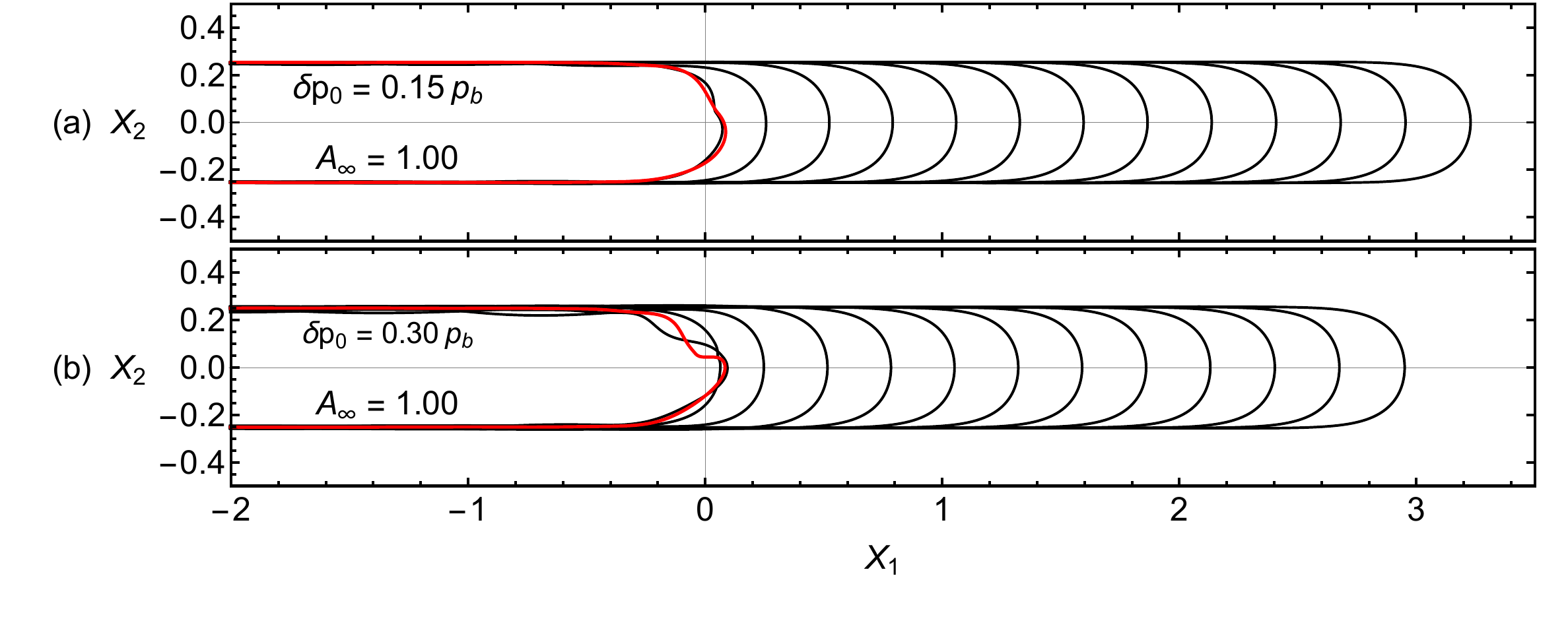}
\caption{Finger propagation for an initially uncollapsed channel, $A_\infty = 1$. Unsteady numerical simulations were initialised with a steady solution at $Ca=0.47$. The displacement of the interfaces is calculated by integrating the frame speed $u_{f}$ in time. The time separation between the interfaces is $\delta t = 0.2$. During the time evolution, the interface is subject to a transient local pressure perturbation in the form of equation (\ref{pressure_perturb}). The amplitude of the perturbation is (a) $\delta p_{0} = 0.15 p_{b}$ and (b) $\delta p_{0} = 0.30 p_{b}$ (b). The interfaces at the time $t_{p}$ are highlighted in red. The deformation on the interface quickly decays and a steadily-propagating symmetric finger is established.}
\label{Time_A1}
\end{figure}

We apply a localised asymmetric perturbation to the
pressure jump across the interface of the form
\begin{equation}
\delta p = - \delta p_{0} e^{-(t/t_{p})^2}
e^{-((x_{2}-y)/\lambda_{p})^2},
\label{pressure_perturb}
\end{equation}
where $\delta p_{0}$ is the amplitude of the perturbation;
$\lambda_{p}=0.035$ is the width of the perturbation; $y=0.005$
is the offset from the centreline; and the time scale $t_{p}=0.015$.
We choose the pressure perturbations to be fractions of $p_{b}$, usually $\delta p_{0} = 0.15 p_{b}$ and $\delta p_{0} = 0.3 p_{b}$, to ensure that we
apply comparable perturbations for different levels of collapse. The
perturbation leads to the formation of a controlled dimple at the
interface as indicated by the finger outlines highlighted in red in figure \ref{Time_A1}, which correspond to the interface shape at time $t_p$.

For $A_\infty = 1$, figure \ref{Time_A1}, the initially
deformed finger rapidly relaxes to the linearly stable, symmetric,
steady state identified in figure \ref{Pb_Ainf} (stable black branch)
for both amplitudes of perturbation, consistent with the experimental
observations discussed in \S \ref{exp_comp}. For $\delta p_0 =0.15p_b$
and $0.30p_b$, the finger reaches the steady state within less than
one-half and one channel widths, respectively, making it easily
observable within the experimental channel. 

\begin{figure}
\center \includegraphics[scale=0.6]{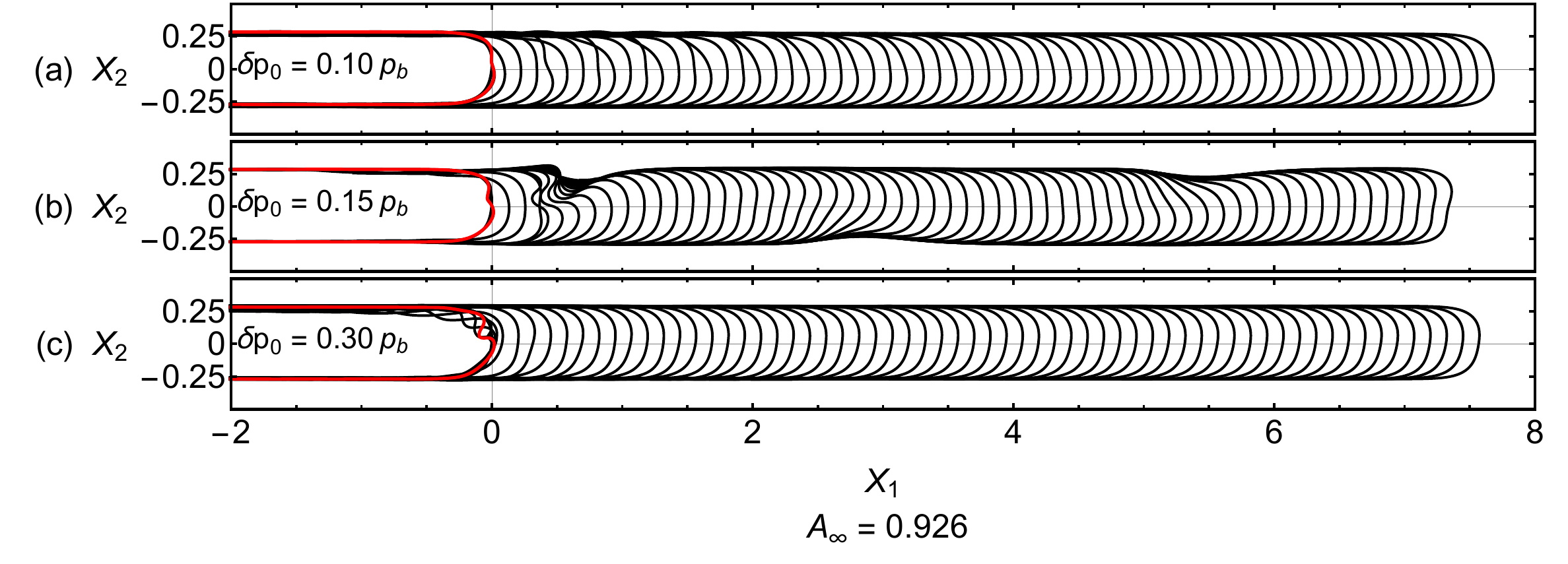}
\center \includegraphics[scale=0.45]{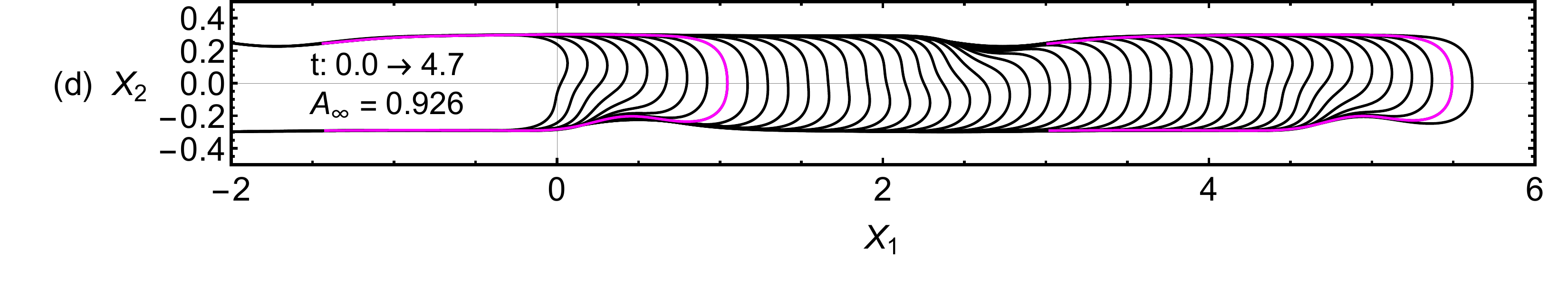}
\caption{
  Finger propagation for an initially slightly-collapsed channel, $A_\infty = 0.926$ and fixed flow rate. Unsteady numerical simulations were initialised with a steady solution at $Ca=0.47$. The displacement of the interfaces is calculated by integrating the frame speed $u_{f}$ in time. The time separation between the interfaces is $\delta t = 0.1$  During the time evolution, the interface is subject to a transient local pressure perturbation given by equation (\ref{pressure_perturb}). The amplitude of the perturbation is (a) $\delta p_{0} = 0.10 p_{b}$; (b) $\delta p_{0} = 0.15 p_{b}$ and (c) $\delta p_{0} = 0.3 p_{b}$. The interfaces at the time $t_{p}$ are highlighted in red. For the smallest-amplitude perturbation the interface 
quickly returns to the linearly stable steadily-propagating asymmetric state. For the intermediate-amplitude perturbation the finger evolves towards a periodic solution, shown in (d). The first and last interfacial positions of one complete oscillation are highlighted in pink and the period $T=3.6$. For the largest-amplitude perturbation, the finger evolves towards the alternative steadily-propagating asymmetric state, in which the asymmetric finger has been reflected about the channel's centreline.}
  \label{time_evolution2}
\end{figure}

Figure \ref{time_evolution2} shows the time evolution of the
finger for $A_{\infty} =  0.926$, after the pitchfork bifurcation $P_{1}$, but before the Hopf bifurcation $H_{1}$, which means that there are two linearly stable steady asymmetric solutions, each being the reflection of the other about the channel's centreline. If one of these steadily-propagating, asymmetric fingers 
is subject to a small pressure perturbation $\delta p_{0} = 0.1 p_{0}$, figure \ref{time_evolution2}(a), then 
the finger initially exhibits small amplitude asymmetric oscillations, but these decay and the finger returns to the steadily-propagating state. The asymmetric oscillations are consistent with the shape of eigenmode associated with the least-stable eigenvalue, but because the perturbation excites a number of eigenmodes the oscillation frequency does not match that of the least-stable eigenmode. In fact, the linear stability analysis shows that there are a large number of near-neutral oscillatory modes for $A_{\infty} < 0.93$ and their presence leads to a non-trivial oscillatory response to general perturbations. 

For a larger perturbation amplitude, $\delta p_{0} = 0.15 p_{0}$, figure \ref{time_evolution2}(b), the finger does not return to the
steadily-propagating state but instead exhibits periodic oscillations. The finger tip advances alternately on either side of the channel and the periodic state has a spatio-temporal ``shift and reflect'' symmetry: it is invariant under reflection about the channel's centreline combined with temporal shift by half a period. A complete period of the final periodic state is shown in figure \ref{time_evolution2}(d). For an even larger perturbation amplitude, $\delta p_{0} = 0.3 p_{0}$, figure \ref{time_evolution2}(c), the finger adopts the alternative steadily-propagating asymmetric state: the final finger shape in figure \ref{time_evolution2}(c) is the same as that in figure \ref{time_evolution2}(a) after reflection about the channel's centreline. Thus there are, at least, three possible stable states at this level of collapse. The periodic state can be continued to values of $A_{\infty}$ above and below $A_{\infty} = 0.926$ by smoothly changing $A_{\infty}$ during the time simulation and waiting until the system settles into a new periodic state. The periodic state appears to persist for increasing values of $A_{\infty}$ until the limit point at $A_{\infty} = 0.927$ with little change in period.



\begin{figure}
\center
\includegraphics[scale=0.6]{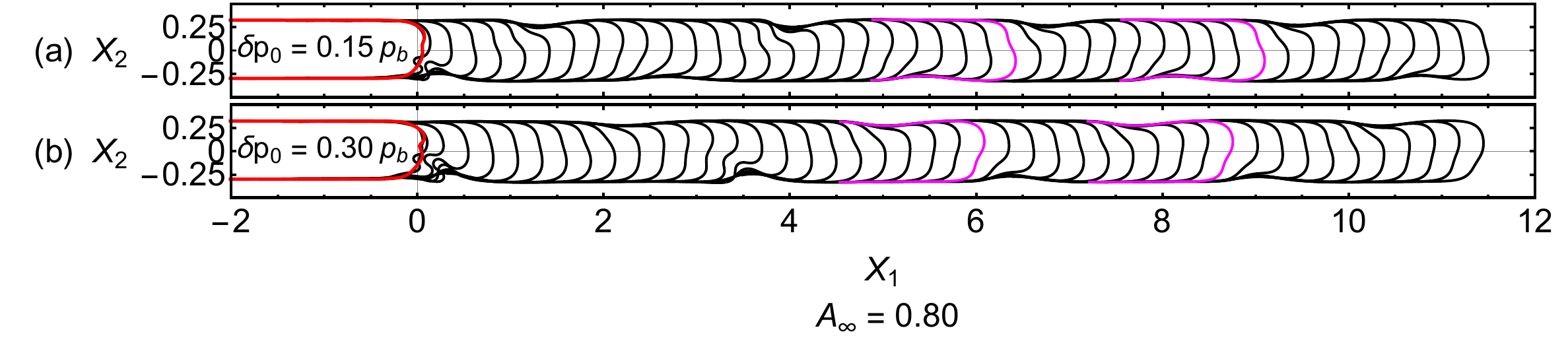}
\caption{Finger propagation for a moderately collapsed channel, $A_\infty =0.8$ Unsteady numerical simulations were initialised with a steady solution at $Ca=0.47$. The displacement of the interfaces is calculated by integrating the instantaneous frame speed $u_{f}$ in time. The time separation between the interfaces is $\delta t = 0.23$. During the time evolution, the interface is subject to a transient local pressure perturbation in the form of equation (\ref{pressure_perturb}). The amplitude of the perturbation is $\delta p_{0} = 0.15 p_{b}$ (a) and $\delta p_{0} = 0.30 p_{b}$ (b). The interfaces at the time $t_{p}$ are highlighted in red and
 the deformation on the interface is advected to the narrower side of the asymmetric finger and decreases in amplitude. The finger tip rapidly develops an oscillatory motion. The first and last interfacial positions of one oscillation are highlighted in pink, The period and wavelength of the oscillation are $T=2.3$ and $L=2.7$ channel widths, respectively. In order to aid the visualization of the oscillations, we are only plotting the interfaces for 1.5 channel widths behind the finger tip.}
\label{Time_A80}
\end{figure}

As the initial level of collapse is reduced further to $A_\infty=0.8$,
figure \ref{Time_A80}, the steadily-propagating asymmetric states become linearly unstable to a complex conjugate pair of eigenvalues through the Hopf bifurcation $H_{1}$. When the asymmetric finger is perturbed the localised dimple initially applied to the
interface increases in length as the finger advances and is advected to the side of the finger, where it decays more rapidly for the smaller perturbation
amplitude. For both perturbation amplitudes, the finger rapidly relaxes to the time-periodic state first observed at $A_{\infty} = 0.926$. These results together with those for $A_{\infty} = 0.926$ indicate that the Hopf bifurcation is subcritical, but the relationship between the associated unstable asymmetric limit cycles and the observed stable symmetric limit cycle was not investigated.

Although there is evidence of
small-amplitude, transient oscillations in the experiments, the length of the observation window available to \cite{Ducloue2017a}, approximately eight channel widths, was insufficient for reliable detection of the large amplitude periodic state.
Nevertheless, the system exhibits a similar response to perturbations: a localised cleft can be seen on the upper side of the snapshot of the experimental finger in figure \ref{Interface}(c) for $A_\infty=0.70$ that is similar to the cleft that develops as the dimple is advected to the side of the finger in the simulatioons. 

\begin{figure}
\center
\includegraphics[scale=0.37]{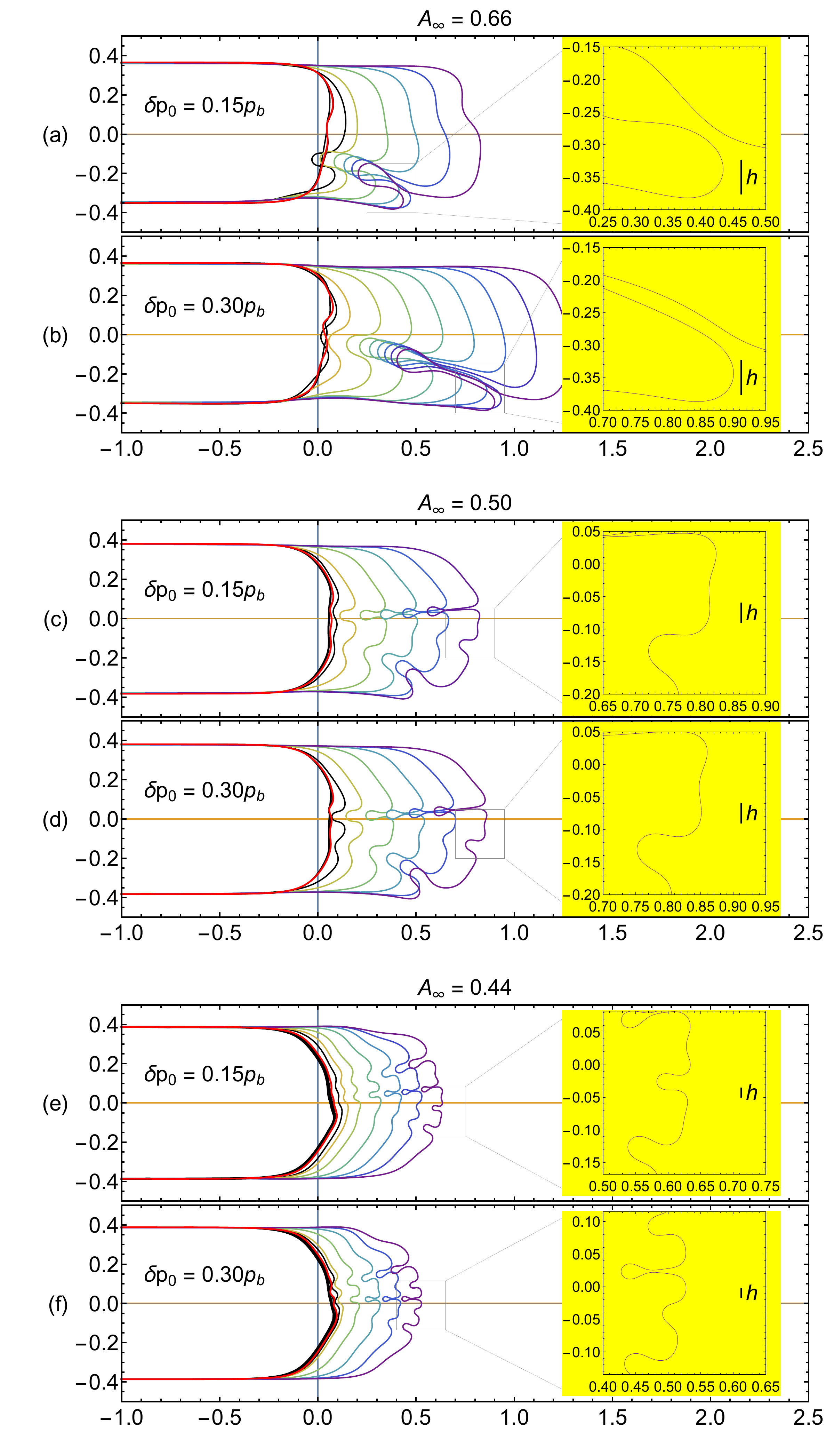}
\caption{Finger propagation for high levels of collapse, $A_{\infty}<A_{\infty}(H_{2})=0.69$. For each value of $A_{\infty}$, unsteady numerical simulations were initialised with a steady solution at $Ca=0.47$. The displacement of the interfaces is calculated by integrating $u_{f}$ in time. The time separation between the interfaces is $\delta t$. During the time evolution, the interface is subject to the same pressure perturbation used in figure \ref{Time_A1}. The amplitude of the perturbation is $\delta p_{0} = 0.15 p_{b}$ (a,c,e) and $\delta p_{0} = 0.30 p_{b}$ (b,d,f). The interfaces at the time $t_{p}$ are highlighted in red. The insets in yellow magnify the regions of the fingering instabilities and include a scale bar to indicate the height, $h$, of the reopening membrane at the centreline of the channel at the $x_{1}$ position of the finger tip. The latter interfaces are plotted in a colour gradient to aid visualization. For $A_{\infty} = 0.66$ in (a) and (b) where $\delta t = 0.15$ and $h = 0.051$, the deformation on the interface is advected to the narrower side of the asymmetric finger and evolves to a deep indentation. For $A_{\infty} = 0.50$ in (c) and (d) where $\delta t = 0.15$ and $h = 0.029$, the initial localized deformation destabilizes the interface and multiple small scale fingers emerge. For $A_{\infty} = 0.44$ in (e) and (f) where $\delta t = 0.10$ and $h = 0.013$, the evolution is similar to (c) and (d) but with an even smaller typical wave-length of the fingering pattern. We stop the time evolution of the fingers presented in this figure when the interface is about to self-intersect.}
\label{Time_evolution_2}
\end{figure}

At $A_\infty = 0.66$, figure \ref{Time_evolution_2}(a,b), the asymmetric
finger is unstable to two oscillatory eigenmodes because the second Hopf
bifurcation point $H_2$ has been crossed. Both amplitudes of
perturbation now result in the formation of a narrower dimple than in figures
\ref{Time_A80}(a,b). The dimple grows in depth as the finger propagates,
while advecting around one side of the finger tip. The larger perturbation creates a deeper cleft.
In both cases, the cleft narrows as it grows and a neck region
is formed. We discontinued the numerical simulations when the two faces of the cleft made contact in the neck region.

The pitchfork point $P_{2}$ has been crossed by $A_\infty =0.5$
(figure \ref{Time_evolution_2}(c,d)) so that the unstable, steady state is
now symmetric (unstable black branch in figure \ref{Pb_Ainf}). The
initial perturbation whose width is on the order of the depth of the
fluid layer now evolves into several clefts. The narrowest cleft is close
to the channel centreline and its size remains close to the length-scale of the
initial perturbation. Moreover, the clefts emerge on both sides of the
centreline in contrast with $A_\infty=0.66$, where a single cleft grew
on one side of the centreline due to the asymmetric nature of the
initial perturbation. They are advected around the tip of the finger
until the numerical simulation had to be discontinued due to contact between
the faces of the narrowest, most centred cleft.  A similar evolution
is found in figure \ref{Time_evolution_2}(e,f) for $A_{\infty}=0.44$,
where the formation of four small clefts result in a pattern of five
small-scale stubby fingers on the propagating front before the
numerical simulation had to be discontinued. This pattern is qualitatively
similar to that observed  experimentally in figure
\ref{Interface}(e). 

Overall, figure \ref{Time_evolution_2} indicates that the widths and depths
of the features that develop on the interface are reduced as the level
of collapse is increased from $A_\infty=0.66$ to $0.44$ in qualitative
agreement with the experimental images shown in figure \ref{Interface}. The observed features form on the
approximate length-scale of the depth of the layer as shown by \cite{Ducloue2017b}
and their characteristic size is reduced as $A_\infty$ decreases
because the depth of the layer in the centre of the channel decreases
accordingly.  In the experiments of \cite{Ducloue2017a}, the
interfacial patterns form continuously at the propagating front and
are advected around the tip of the finger where they decay. Because
our numerical simulations had to be discontinued when the interface self-intersects within the clefts, we were unable to determine whether the tip
instabilities driven by the initially imposed perturbation would
eventually decay or whether they would be sustained as in the
experiments.  The linear stability analysis showed, however, that there are a large number of near-neutral oscillatory modes that could interact non-linearly to yield non-trivial transients. Thus, although the unsteady simulations of the numerical model exhibit many of the same features as the experiments, we do not yet have a detailed understanding of how the complex solution structure leads to the development of the observed small-scale fingers. 

\section{Discussion and conclusion}
\label{Conclusion}

 In this paper we have presented a depth-averaged model to describe the propagation of an air finger into a collapsed elasto-rigid channel filled with viscous liquid and driven at constant volume flux. We find that the model is in excellent qualitative and quantitative agreement with the experiments of \cite{Ducloue2017a}. The model predicts a non-trivial solution structure with multiple co-existing steady and oscillatory modes of propagation for the same level of initial collapse and finger propagation speed.

 In line with the experiments, the complexity of the solution structure increases with the level of collapse \citep{Ducloue2017a} for a fixed finger propagation speed. At low levels of collapse, the interface propagates steadily with a morphology similar to  a Saffman--Taylor finger in a rigid channel. The model predicts the existence of an alternative, unstable, double-tipped, steadily-propagating finger analogous to the first Romero--Vanden-Broeck solution in rigid channel \citep{Romero1982,VandenBroeck1983,mccue2015}, which requires a greater finger pressure to propagate at the same speed as the Saffman--Taylor-like finger. 

 For $0.93 >  A_\infty > 0.61$ there are two steadily-propagating fingers each with an asymmetric tip arising through a symmetry-breaking bifurcation from the Saffman--Taylor solution and related by reflection about the centreline of the channel. The asymmetric fingers are only stable for a very small range of $A_{\infty}$ and lose stability to asymmetric oscillatory modes through a Hopf bifurcation at $A_{\infty} = 0.922$. The close proximity of the limit point, pitchfork and Hopf bifurcations suggests that they may arise from the perturbation of a bifurcation of higher co-dimension. An analogous solution structure has been found in two-phase flow through a uniformly curved rigid tube \citep{HazelHeilEtAl2012} in which case it could be shown to arise directly from a perturbed  fold-Hopf bifurcation \citep{Kuznetsov1998}.

 Time-dependent simulations at fixed flow rate showed that for values of $A_{\infty}$ lower than $0.922$ if the unstable steadily-propagating finger is perturbed it will eventually settle on an oscillatory mode of propagation in which the finger tip meanders from side. Irregular meandering of the finger tip has been seen in large-aspect-ratio, rigid, Hele-Shaw cells \citep{Moore2003}. We have found that this stable oscillatory mode persists for value of $A_{\infty} > 0.922$ indicating that the Hopf bifurcation is subcritical. We also find that the oscillatory mode can be reached by applying a suitable nonlinear perturbation to the stable steadily-propagating asymmetric finger when $0.927 > A_{\infty} > 0.922$.

\begin{figure}
\center
\includegraphics[scale=0.7]{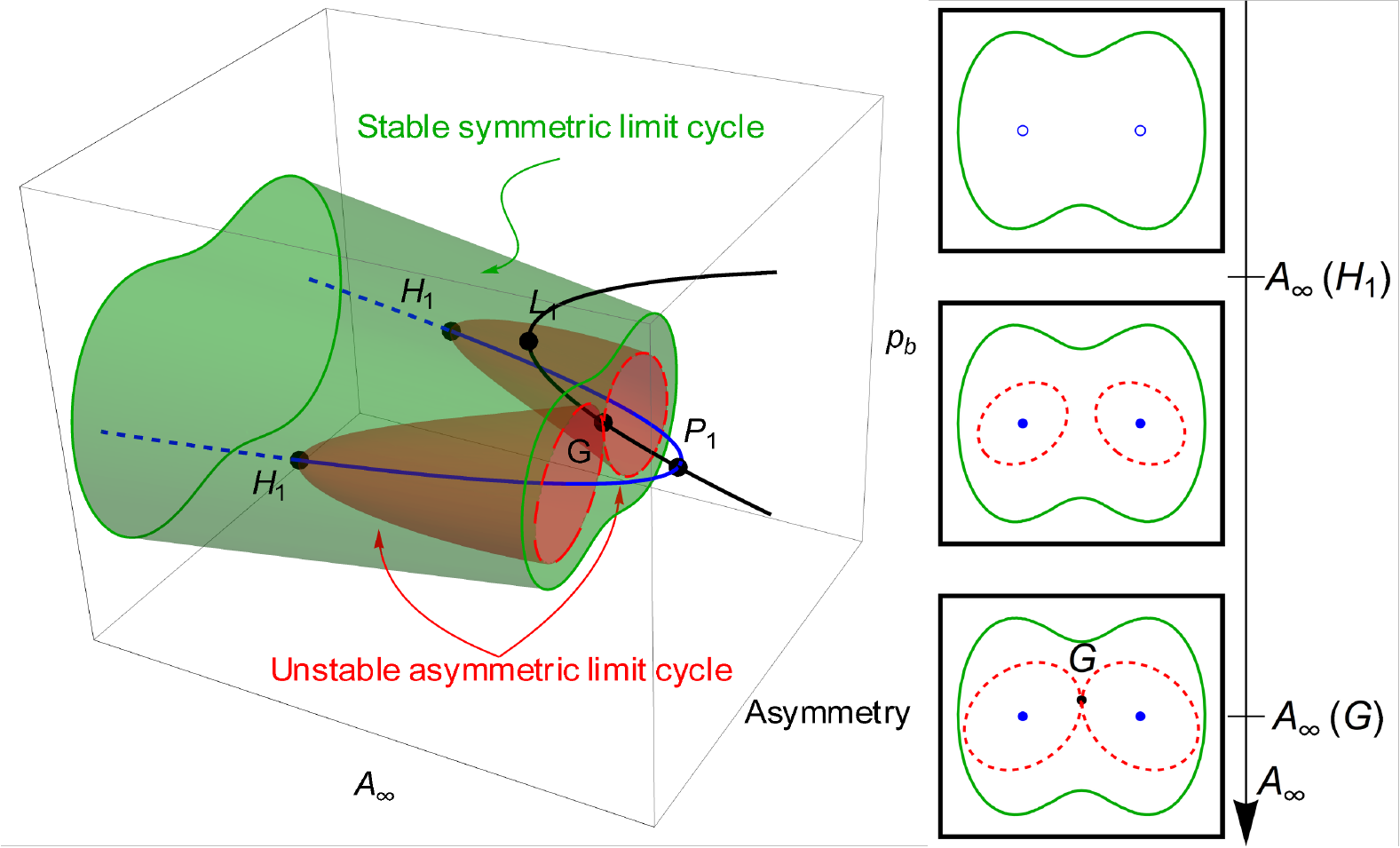}
\caption{A sketch of the proposed bifurcation scenario in the region $0.9 < A_{\infty} < 0.95$ at fixed $Ca=0.47$. A symmetric steadily-propagating finger state (black line)  undergoes a symmetry-breaking pitchfork bifurcation at $A_{\infty} = 0.93$ followed by a limit point at $A_{\infty} = 0.927$. The two steadily-propagating asymmetric states that arise from the pitchfork bifurcation (blue line) each undergo a subcritical Hopf bifurcation at $A_{\infty} = 0.922$. Two unstable limit cycles emanate from the Hopf bifurcation on each asymmetric branch for $A_{\infty} > 0.922$. We conjecture that these unstable limit cycles become a symmetric limit cycle via a gluing bifurcation on the symmetric branch in the region between the limit point and the pitchfork bifurcation, $0.927 > A_{\infty} > 0.93$. The symmetric limit cycle is likely to be stabilized through a limit point (not shown).}
\label{fig:gluing_sketch}
\end{figure}

 The observed oscillations have a spatio-temporal symmetry being identical under reflection about the channel's centreline after a time-shift of half a period. Hence, the periodic state cannot arise directly from the Hopf bifurcation because at the bifurcation two unstable, asymmetric limit cycles will emanate from the two asymmetric steadily-propagating solution branches. Instead, the observed periodic state must be a consequence of another bifurcation that we have not identified. The simplest possibility is that the two asymmetric limit cycles merge to create the symmetric limit cycle, a process known as a gluing bifurcation \citep{Kuznetsov1998}, see figure \ref{fig:gluing_sketch}.  An unstable symmetric steady state is involved in a standard gluing bifurcation, which means that the symmetric periodic solution cannot be created until $A_{\infty}$ is greater that the limit point $L_{1}$ on the symmetric branch. The existence of a stable symmetric limit cycle for $A_{\infty} < 0.927$ suggests that a limit point of periodic states must also exist either for the symmetric cycle after gluing or for the two asymmetric cycles before gluing. A sketch of the former scenario is shown in figure \ref{fig:gluing_sketch} and is consistent with a perturbed Takens-Bogdanov bifurcation with underlying $\mathbbm{Z}_{2}$ symmetry, see for example figure 2 in \cite{rucklidge1993}. This co-dimension two bifurcation
has been identified as the organising centre for complex dynamics in other scenarios, such as double-diffusive convection \citep{knobloch_proctor_1981} and magnetoconvection \citep{rucklidge1993}. 

At increased levels of collapse, the asymmetric fingers are further destabilised through a second Hopf bifurcation at $A_{\infty} = 0.69$ and rather than settling into the periodic state the interface exhibits a more complex response to perturbations. In the least collapsed channels, a single cleft develops in the interface, which resembles the early stages of tip-splitting instabilities in rigid Hele-Shaw cells. As the level of collapse increases further, the number of clefts increases and the morphology resembles a small-scale fingering instability of the tip. Instability of the finger tip to small-scale fingers was observed in the experiments by \cite{Ducloue2017a} for $A_{\infty} < 0.7$. The typical length-scale of the smaller fingers decreases with increasing levels of collapse in both the experiments and predictions of the model. The length-scale of the small-scale fingers is comparable to the height of the channel, which means that the interface configuration violates one of the assumptions of the model: variations in the transverse direction should occur over a greater length-scale than the channel height. In the model, the interface eventually self-intersects forcing us to terminate the computations at these points. Hence, it is not possible to determine whether these small-scale fingering patterns are transient or self-sustaining. A further complication is the presence of a large number of near-neutral oscillatory modes in the model that are likely to lead to complex transient dynamics. Near contact of the interface was not observed in experiments, implying that effects not included in the model may prevent self-intersection and providing further evidence that the results of the model in which the interface has transverse variations over small length-scales should be treated with caution.

We chose to apply the liquid-film corrections using an
effective capillary number based on the average interface velocity instead of
using a local capillary number based on the velocity at each point of
the interface. For a steadily-propagating interface, every point must move
with the same velocity so all local capillary numbers will be the
same. In time-dependent simulations,
however, particularly for the unstable fingers, the velocity can vary
significantly along the interface leading to variations in the local
capillary number. In physical terms, basing the liquid-film corrections on the average velocity means that in the model the
total thickness of the liquid films relative to the local channel
height is assumed to be constant and that there are no pressure gradients within the films. In reality the liquid films will vary in thickness around the finger. The good quantitative agreement between the model predictions and experimental data suggests that these fine details do not influence the finger width far behind the tip nor the finger pressure. The liquid-film modelling assumptions are likely to have a significant impact on the development of the small-scale fingers, however. 
 
As far as we are aware all previous studies of the influence of liquid-film corrections in Hele-Shaw cells have been for cases in which there is only one solution. In these cases the purpose of the correction is to achieve improved quantitative agreement between the model and experiments \citep{Tabelin1986}, but the qualitative features are unchanged.  In the present study, we have found a non-trivial solution structure with multiple co-existing states. If liquid-film corrections are not applied in our system then the results are both qualitatively and quantitatively wrong: the number and nature of the solutions changes.

We conclude that the relatively simple depth-averaged model appears to capture the majority of the features observed in experiments and, moreover, that the steadily-propagating solutions present in the depth-averaged model of rigid Hele-Shaw channels are also present in the elastic-walled channel. The presence of the elastic wall can lead to interaction between solution branches that are isolated in the rigid channel, altering their stability and leading to complex dynamics in elasto-rigid channels at higher levels of initial collapse.


\section*{Acknowledgements}

This work was supported via EPSRC grants EP/J007927/1 and EP/P026044/1. The preliminary model development was supported by the Leverhulme Trust under grant number RPG-2014-081.

\section*{Declaration of interests}

The authors report no conflict of interest. 

\appendix

\bibliography{ms} \bibliographystyle{jfm}

\end{document}